\documentclass{lmcs} 
\pdfoutput=1

\usepackage{lastpage}
\lmcsdoi{21}{4}{31}
\lmcsheading{}{\pageref{LastPage}}{}{}%
{Mar.~27,~2025}{Dec.~30,~2025}{}

\usepackage{hyperref}

\usepackage{xcolor}
\usepackage{amsfonts}
\usepackage{amssymb}
\usepackage{amsthm}
\usepackage{stmaryrd}
\usepackage{amsmath}
\usepackage[disable]{todonotes}
\usepackage{shuffle}
\usepackage{mathtools}
\usepackage{multirow}
\usepackage{makecell}
\usepackage{xspace}
\usepackage{tikz}
\usepackage{booktabs}
\usepackage[capitalize]{cleveref}
\usepackage[utf8]{inputenc}

\newcommand{\tup}[1]{\langle #1 \rangle}
\newcommand{\sem}[1]{\llbracket #1 \rrbracket}
\renewcommand{\vec}[1]{\mathbf{#1}}

\newcommand{\cA}{\mathcal{A}}
\newcommand{\cB}{\mathcal{B}}
\newcommand{\calC}{\mathcal{C}} 

\newcommand{\jL}{\mathfrak{J}}

\newcommand{\regL}{\mathfrak{L}}

\newcommand{\bbN}{\mathbb{N}}

\newcommand{\bbZ}{\mathbb{Z}}

\newcommand{\NFA}{\mbox{\rm NFA}\xspace}

\newcommand{\NFAs}{\mbox{\rm NFAs}\xspace}
\newcommand{\JFA}{\mbox{\rm JFA}\xspace}
\newcommand{\JFAs}{\mbox{\rm JFAs}\xspace}

\newcommand{\simkw}[1][k]{\sim_{k\boxplus}}

\newcommand{\Abs}{\textsc{abs}}
\newcommand{\Ham}{\textsc{ham}}
\newcommand{\Rev}{\textsc{rev}}
\newcommand{\Max}{\textsc{max}}
\newcommand{\Sem}{\textsc{sem}}

\newcommand{\numrev}{\#_{\Rev}}
\newcommand{\turnindices}{\mathtt{Turn}} 

\newcommand{\range}{\mathtt{Range}}
\newcommand{\numslots}{{2k+2}}
\newcommand{\activate}{\textsc{activate}}
\newcommand{\deactivate}{\textsc{deactivate}}
\newcommand{\none}{\textsc{none}}
\newcommand{\numcross}{\#_\times}

\newcommand{\cxNPc}{\textsf{NP-complete}\xspace}
\newcommand{\cxNP}{\textsf{NP}\xspace}
\newcommand{\cxNPh}{\textsf{NP-hard}\xspace}
\newcommand{\cxPSPACE}{\textsf{PSPACE}\xspace}
\newcommand{\cxPSPACEc}{\textsf{PSPACE-complete}\xspace}
\newcommand{\cxPSPACEh}{\textsf{PSPACE-hard}\xspace}
\newcommand{\cxEXPSPACE}{\textsf{EXPSPACE}\xspace}
\newcommand{\cxEXPEXPSPACE}{\textsf{2-EXPSPACE}\xspace}

\newcommand{\cxsPSPACEc}{\textsf{PSPACE-c}\xspace}
\newcommand{\cxsPSPACEh}{\textsf{PSPACE-h}\xspace}

\newcommand{\shtodo}[1]{\todo[inline,color=cyan!25,size=\small]{SH: #1}}

\newcommand{\probkBoundedness}{\textsc{Univ-$k$-Bnd}\xspace}
\newcommand{\probParamBoundedness}{\textsc{Univ-Param-Bnd}\xspace}

\newcommand{\probMembership}{\textsc{Membership}\xspace}

\newcommand{\probUnivBoundedness}[1]{\textsc{Univ-#1-Bnd}\xspace}

\newcommand{\probDomkBoundedness}{\textsc{$k$-Bnd}\xspace}
\newcommand{\probDomParamBoundedness}{\textsc{Param-Bnd}\xspace}

\usetikzlibrary{automata, arrows.meta, positioning, decorations.pathreplacing}

\crefname{thm}{Theorem}{Theorems}
\crefname{lem}{Lemma}{Lemmas}
\crefname{rem}{Remark}{Remarks}
\crefname{exa}{Example}{Examples}
\crefname{cor}{Corollary}{Corollaries}
\crefname{defi}{Definition}{Definitions}
\crefname{obs}{Observation}{Observations}

\begin{document}

\title{Quantitative Semantics for Jumping Automata}
\thanks{This research was supported by the ISRAEL SCIENCE FOUNDATION (grant No. 989/22). We also thank Micha\"el Cadilhac for fruitful discussions about this submission.\\
This paper is an extended version of~\cite{almagorjumping}.
}	

\author[S.~Almagor]{Shaull Almagor\lmcsorcid{0000-0001-9021-1175}}
\author[N.~Dafni]{Neta Dafni\lmcsorcid{0000-0002-2515-2014}}
\author[I.~Salgado]{Ishai Salgado}

\address{Technion, Israel}	
\email{shaull@technion.ac.il, netad@campus.technion.ac.il, ishaisalgado@campus.technion.ac.il}  






\begin{abstract}
Jumping automata are finite automata that read their input in a non-sequential manner, by allowing a reading head to ``jump'' between positions on the input, consuming a permutation of the input word.
We argue that allowing the head to jump should incur some cost. To this end, we propose four quantitative semantics for jumping automata, whereby the jumps of the head in an accepting run define the cost of the run. The four semantics correspond to different interpretations of jumps: the \emph{absolute distance} semantics counts the distance the head jumps, the \emph{reversal} semantics counts the number of times the head changes direction, the \emph{Hamming distance} measures the number of letter-swaps the run makes, and the \emph{maximum jump} semantics counts the maximal distance the head jumps in a single step, 

We study these measures, with the main focus being the \emph{boundedness problem}: given a jumping automaton, decide whether its (quantitative) language is bounded by some given number $k$. We establish the decidability and complexity for this problem under several variants.
\end{abstract}

\maketitle

\section{Introduction}
\label{sec:intro}
Traditional automata read their input sequentially. Indeed, this is the case for most state-based computational models.
In some settings, however, we wish to abstract away the order of the input letters. 
For example, when the input represents available resources, and we only wish to reason about their \emph{quantity}. From a more language-theoretic perspective, this amounts to looking at the \emph{commutative closure} of languages, a.k.a. their \emph{Parikh image}. 
To capture this notion in a computation model, \emph{Jumping Automata (\JFAs)} were introduced in~\cite{meduna2012jumping}. A jumping automaton may read its input in a non-sequential manner, jumping from letter to letter, as long as every letter is read exactly once. Several works have studied the algorithmic properties and expressive power of these automata~\cite{fernau2015jumping,fernau2017characterization,vorel2018basic,fazekas2021two,lavado2014operational,almagor2023jumping}. 

While \JFAs are an attractive and simple model, they present a shortcoming when thought of as model for systems, namely that the abstraction of the order may be too coarse. 
More precisely, the movement of the head can be thought of as a physical process of accessing the input storage of the \JFA. Then, sequential access is the most basic form of access and can be considered ``cheap'', but allowing the head to jump around is physically more difficult and therefore should incur a cost. 

To address this, we present four \emph{quantitative semantics} for \JFAs, whereby a \JFA represents a function from words to costs, which captures how expensive it is to accept a given word with respect to the head jumps. The different semantics capture different aspects of the cost of jumps, as follows.

Consider a \JFA $\cA$ and a word $w$, and let $\rho$ be an accepting run of $\cA$ on $w$. The run $\rho$ specifies the sequence of states and indices visited in $w$. We first define the cost of individual runs.
\begin{itemize}
    \item In the \emph{Absolute Distance} semantics ($\Abs$), the cost of $\rho$ is the sum of the lengths of jumps it makes.
    \item In the \emph{Reversal} semantics ($\Rev$), the cost of $\rho$ is the number of times the reading head changes its direction (i.e., moving from right to left or from left to right).
    \item In the \emph{Hamming} semantics ($\Ham$), we consider the word $w'$ induced by $\rho$, i.e., the order of letters that $\rho$ reads. Then, the cost of $\rho$ is the number of letters where $w'$ differs from $w$.
    \item In the \emph{Maximum Jump} semantics ($\Max$), the cost of $\rho$ is the maximal length of a jump that it makes.
\end{itemize}
We then define the cost of the word $w$ in $\cA$ according to each semantics, by taking the run that minimizes the cost.

Thus, we lift \JFAs from a Boolean automata model to the rich setting of \emph{quantitative} models~\cite{boker2021quantitative,almagor2011max,chatterjee2010quantitative,droste2009handbook}. Unlike other quantitative automata, however, the semantics in this setting arise naturally from the model, without an external domain.
Moreover, the definitions are naturally motivated by different types of memory access, as we now demonstrate. First, consider a system whose memory is laid out in an array (i.e., a tape), with a reading head that can move along the tape. Moving the head requires some energy, and therefore the total energy spent reading the input corresponds to the $\Abs$ semantics. 
It may be the case, however, that moving the head over several cells requires some mechanical change in the way the memory works, but that once a jump over $k$ cells has been implemented, its cost is no more than any other jump. The, the cost measures the largest jump that needs to be taken. This is captured by the $\Max$ semantics.

Next, consider a system whose memory is a spinning disk (or a sliding tape), so that the head stays in place and the movement is of the memory medium. Then, it is cheap to continue spinning in the same direction\footnote{We assume that the head does not return back to the start by continuing the spin, but rather reaches some end.}, and the main cost is in reversing the direction, which requires stopping and reversing a motor. Then, the $\Rev$ semantics best captures the cost.

Finally, consider a system that reads its input sequentially, but is allowed to edit its input by replacing one letter with another, such that at the end the obtained word is a permutation of the original word. This is akin to \emph{edit-distance automata}~\cite{mohri2002edit,fisman2023normalized} under a restriction of maintaining the amount of resources. Then, the minimal edits required correspond to the $\Ham$ semantics. 

\begin{exa}
\label{xmp:semantics ab simple}
Consider a (standard) \NFA $\cA$ for the language given by the regular expression $(ab)^*$ (the concrete automaton chosen is irrelevant, see~\cref{rmk:semantics semantics}).
As a \JFA, $\cA$ accepts a word $w$ if and only if $w$ has an equal number of $a$'s and $b$'s.
    To illustrate the different semantics, consider the words $w_1=ababbaab$ and $w_2=ababbaba$, obtained from $(ab)^4$ by flipping the third $ab$ pair ($w_1$) and the third and fourth pairs ($w_2$). As we define in~\cref{sec:quantitative semantics}, we think of runs of the \JFA $\cA$ as if the input is given between end markers at indices $0$ and $n+1$, and the jumping run must start at $0$ and end in $n+1$, respectively.
    \begin{itemize}
        \item In the $\Abs$ semantics, the cost of $w_1$, denoted $\cA_\Abs(w_1)$, is $2$: the indices read by the head are $0,1,2,3,4,6,5,7,8,9$, so there are two jumps of cost $1$: from $4$ to $6$ and from $5$ to $7$. Similarly, we have $\cA_\Abs(w_2)=4$, e.g., by the sequence $0,1,2,3,4,6,5,8,7,9$, which has two jumps of cost $1$ ($4$ to $6$ and $7$ to $9$), and one jump of cost $2$ ($5$ to $8$).  (formally, we need to prove that there is no better run, but this is not hard to see).
        \item In the $\Rev$ semantics, we have $\cA_\Rev(w_1)=2$ by the same sequence of indices as above, as the head performs two ``turns'', one at index $6$ (from $\to$ to $\leftarrow$) and one at $5$ (from $\leftarrow$ to $\to$). 
        Here, however, we also have $\cA_\Rev(w_2)=2$, using the sequence $0,1,2,3,4,6,8,7,5,9$, whose turning points are $8$ and $5$. 
        \item In the $\Ham$ semantics we have $\cA_\Ham(w_1)=2$ and $\cA_\Ham(w_2)=4$, since we must change the letters in all the flipped pairs for the words to be accepted.
        \item In the $\Max$ semantics, we have $\cA_\Max(w_1)=1$, by the same sequence of indices as in the $\Abs$ semantics, as there are two jumps of cost $1$ and so the maximum is $1$. For $w_2$ we have $\cA_\Max(w_1)=2$, also by the same sequence as the $\Abs$ semantics. (again, we need to show that the sequence is optimal).
    \end{itemize}
\end{exa}
\begin{exa}
\label{xmp:semantics anbn unbounded}
Consider now an \NFA $\cB$ for the language given by the regular expression $a^*b^*$. Note that as a \JFA, $\cB$ accepts $\{a,b\}^*$, since every word in $\{a,b\}^*$ can be reordered to the form $a^*b^*$.  

Observe that in the $\Rev$ semantics, for every word $w$ we have $\cB_\Rev(w)\le  2$, since at the worst case $\cB$ makes one left-to-right pass to read all the $a$'s, then a right-to-left pass to read all the $b$'s, and then jump to the right end marker, and thus it has two turning points. In particular, $\cB_\Rev$ is bounded. 

However, in the $\Abs$, $\Ham$ and $\Max$ semantics, the costs can become unbounded. Indeed, in order to accept words of the form $b^na^n$, in the $\Abs$ and $\Max$ semantics the head must first jump over all the $b^n$, incurring a high cost, and for the $\Ham$ semantics, all the letters must be changed, again incurring a high cost. 
\end{exa}

\paragraph*{Related work}
Jumping automata were introduced in~\cite{meduna2012jumping}. We remark that~\cite{meduna2012jumping} contains some erroneous proofs (e.g., closure under intersection and complement, also pointed out in~\cite{fernau2017characterization}). The works in~\cite{fernau2015jumping,fernau2017characterization} establish several expressiveness results on jumping automata, as well as some complexity results. In~\cite{vorel2018basic} many additional closure properties are established. An extension of jumping automata with a two-way tape was studied in~\cite{fazekas2021two}, and jumping automata over infinite words were studied by the first author in~\cite{almagor2023jumping}.

When viewed as the commutative image of a language, jumping automata are closely related to Parikh Automata~\cite{klaedtke2003monadic,cadilhac2012bounded,cadilhac2012affine,guha2022parikh}, which read their input and accept if a certain Parikh image relating to the run belongs to a given semilinear set (indeed, we utilize the latter in our proofs). 
Another related model is that of symmetric transducers -- automata equipped with outputs, such that permutations in the input correspond to permutations in the output. These were studied in~\cite{almagor2020process} in a jumping-flavour, and in~\cite{nassar2022simulation} in a quantitative $k$-window flavour.

More broadly, quantitative semantics have received much attention in the past two decades, with many motivations and different flavors of technicalities. We refer the reader to~\cite{boker2021quantitative,droste2009handbook} and the references therein.

\paragraph*{Contribution and paper organization} 
Our contribution consists of the introduction of the four jumping semantics, and the study of decision problems pertaining to them (defined in \cref{sec:quantitative semantics}). Our main focus is the boundedness problem: given a \JFA $\cA$, decide whether the function described by it under each of the semantics is bounded by some constant $k$. We establish the decidability of this problem for all the semantics, and consider the complexity of some fragments. More precisely, we consider several variants of boundedness, depending on whether the bound is a fixed constant, or an input to the problem, and on whether the jumping language of $\cA$ is universal. Our complexity results are summarized in~\cref{tab:results}.

Our paper is organized as follows: the preliminaries and definitions are given in~\cref{sec:prelim,sec:quantitative semantics}. Then, each of~\cref{sec:absolute distance,sec:reversal,sec:hamming,sec:maxjump} studies one of the semantics, and follows the same structure: we initially establish that the membership problem for the semantics is $\cxNPc$. Then we characterize the set of words whose cost is at most $k$ using a construction of an \NFA. For the $\Abs,\Rev$, and $\Ham$ semantics, these constructions differ according to the semantics, and involve some nice tricks with automata, but are technically not hard to understand. 
In contrast, the construction for $\Max$ is significantly more involved (see \cref{sec:maxjump}).
We note that these constructions are preceded by crucial observations regarding the semantics, which allow us to establish their correctness. 
Next, in~\cref{sec:interplay} we give a complete picture of the interplay between the different semantics (using some of the results established beforehand). In particular, we show the difficulty in working with the $\Max$ semantics (see \cref{xmp:bounded max unbounded rev}).
Finally, in~\cref{sec:discussion} we discuss some exciting open problems.

\begin{table}[htbp]
    \centering
    \begin{tabular}{cccccc}
        \toprule
        & \multirow{2}{*}{\probDomkBoundedness} 
        & \multirow{2}{*}{\probDomParamBoundedness} 
        & \multirow{2}{*}{\probkBoundedness} 
        & \multicolumn{2}{c}{\probParamBoundedness} \\ 
        \cmidrule(lr){5-6}
        & & & & unary & binary \\ 
        \midrule
        $\Abs$  & \makecell{Decidable\\ $\cxsPSPACEh$} 
                & \makecell{Decidable\\ $\cxsPSPACEh$} 
                & $\cxsPSPACEc$  
                & \makecell{$\cxEXPSPACE$ \\ $\cxsPSPACEh$} 
                & \makecell{$\cxEXPEXPSPACE$ \\$\cxsPSPACEh$} \\ 
        $\Rev$  & \makecell{Decidable\\ $\cxsPSPACEh$} 
                & \makecell{Decidable\\ $\cxsPSPACEh$} 
                & $\cxsPSPACEc$  
                & \makecell{$\cxEXPSPACE$ \\ $\cxsPSPACEh$} 
                & \makecell{$\cxEXPEXPSPACE$ \\$\cxsPSPACEh$} \\ 
        $\Ham$  & \makecell{Decidable\\ $\cxsPSPACEh$} 
                & \makecell{Decidable\\ $\cxsPSPACEh$} 
                & $\cxsPSPACEc$  
                & $\cxsPSPACEc$ 
                & \makecell{$\cxEXPSPACE$ \\$\cxsPSPACEh$} \\ 
        $\Max^\dagger$  & \makecell{Decidable\\ $\cxsPSPACEh$} 
                & \makecell{Decidable\\ $\cxsPSPACEh$} 
                & $\cxsPSPACEc$  
                & \makecell{$\cxEXPSPACE$ \\ $\cxsPSPACEh$} 
                & \makecell{$\cxEXPEXPSPACE$ \\$\cxsPSPACEh$} \\ 
        \bottomrule
    \end{tabular}
    \caption{Complexity results of the various boundedness problems for the four semantics. The complexity of membership is $\cxNPc$ for all the semantics. The ``Decidable'' entries depend on the complexity of the containment problem for Parikh Automata. ($\dagger$) -- for the $\Max$ semantics, PSPACE-hardness is only for $k=0$.}
    \label{tab:results}
\end{table}

\section{Preliminaries and Definitions}
\label{sec:prelim}
For a finite alphabet $\Sigma$ we denote by $\Sigma^*$ the set of finite words over $\Sigma$. 
For $w\in \Sigma^*$ we denote its letters by $w=w_1\cdots w_n$, and its length by $|w|=n$.
In the following, when discussing sets of numbers, we define $\min\emptyset=\infty$.
\paragraph*{Automata}
A \emph{nondeterministic finite automaton (\NFA)} is a 5-tuple $\cA=\tup{\Sigma,Q,\delta,Q_0,\alpha}$ where $\Sigma$ is a finite alphabet, $Q$ is a finite set of states, $\delta:Q\times\Sigma\to 2^Q$ is a nondeterministic transition function, $Q_0\subseteq Q$ is a set of initial states, and $\alpha\subseteq Q$ is a set of accepting states. 
A \emph{run} of $\cA$ on  a word $w=w_1 w_2 \ldots w_n$ is a sequence $\rho=q_0,q_1,\ldots,q_n$ such that $q_0\in Q_0$ and for every $0\le i<n$ it holds that $q_{i+1}\in \delta(q_i,w_{i+1})$.
The run $\rho$ is \emph{accepting} if $q_{n}\in \alpha$. A word $w$ is \emph{accepted} by $\cA$ if there exists an accepting run of $\cA$ on $w$.
The \emph{language} of $\cA$, denoted $\regL(\cA)$, is the set of words accepted by $\cA$.

\paragraph*{Permutations} 
Let $n\in \bbN$. The \emph{permutation group} $S_n$ is the set of bijections (i.e. \emph{permutations}) from $\{1,...,n\}$ to itself. 
$S_n$ forms a group with the function-composition operation and the identity permutation as a neutral element. 
Given a word $w=w_1\cdots w_n$ and a permutation $\pi\in S_n$, we define
$
\pi(w)=w_{\pi(1)}\cdots w_{\pi(n)}
$.
For example, if $w=abcd$ and $\pi=\begin{pmatrix}
    1 & 2& 3 & 4\\
    3 & 4& 2& 1
\end{pmatrix}$ then $\pi(w)=cdba$.
We usually denote permutations in \emph{one-line form}, e.g., $\pi$ is  represented as $(3,4,2,1)$. 
We say that a word $y$ is a \emph{permutation} of $x$, and we write $x\sim y$ if there exists a permutation $\pi\in S_{|x|}$ such that $\pi(x)=y$.

\paragraph*{Jumping Automata}
A jumping automaton is syntactically identical to an \NFA, with the semantic difference that it has a reading head that can ``jump'' between indices of the input word. An equivalent view is that a jumping automaton reads a (nondeterministically chosen) permutation of the input word.

Formally, consider an \NFA $\cA$. We view $\cA$ as a \emph{jumping finite automaton} (\JFA) by defining its \emph{jumping language}
$\jL(\cA)=\{w\in \Sigma^*\mid \exists u\in \Sigma^*.\ w\sim u\wedge u\in\regL(\cA)\}$. 

Since our aim is to reason about the manner with which the head of a  \JFA jumps, we introduce a notion to track the head along a run. 
Consider a word  $w$ of length $n$ and a \JFA $\cA$. 
A \emph{jump sequence} is a vector $\vec{a} = (a_0, a_1, a_2, \ldots, a_n, a_{n+1})$ where $a_0=0$, $a_{n+1}=n+1$ and $(a_1, a_2, \ldots, a_n)\in S_n$. We denote by $J_n$ the set of all jump sequences of size $n+2$. 

Intuitively, a jump sequence $\vec{a}$ represents the order in which a \JFA visits a given word of length $n$. First it visits the letter at index $a_1$, then the letter at index $a_2$ and so on. To capture this, we define $w_{\vec{a}}=w_{a_1}w_{a_2}\cdots w_{a_n}$.
Observe that jump sequences enforce that the head starts at position 0 and ends at position $n+1$, which can be thought of as left and right markers, as is common in e.g., two-way automata.

An alternative view of jumping automata is via \emph{Parikh Automata (PA)}~\cite{klaedtke2003monadic,cadilhac2012affine}. The standard definition of PA is an automaton whose acceptance condition includes a semilinear set over the transitions. To simplify things, and to avoid defining unnecessary concepts (e.g., semilinear sets), for our purposes, a PA is a pair $(\cA,\calC)$ where $\cA$ is an NFA over alphabet $\Sigma$, and $\calC$ is a \JFA over $\Sigma$. Then, the PA $(\cA,\calC)$ accepts a word $w$ if $w\in \regL(\cA)\cap \jL(\calC)$. Note that when $\regL(\cA)=\Sigma^*$, then the PA coincides with $\jL(\calC)$. Our usage of PA is to obtain the decidability of certain problems. Specifically, from~\cite{klaedtke2003monadic} we have that emptiness of PA is decidable.

\section{Quantitative Semantics for \JFAs}
\label{sec:quantitative semantics}
In this section we present and demonstrate the three quantitative semantics for \JFAs. We then define the relevant decision problems, and lay down some general outlines to solving them, which are used in later sections.
For the remainder of the section fix a \JFA $\cA=\tup{\Sigma,Q,\delta,Q_0,\alpha}$.

\subsection{The Semantics}

\paragraph*{The Absolute-Distance Semantics}
In the absolute-distance semantics, the cost of a run (given as a jump sequence) is the sum of the sizes of the jumps made by the head. Since we want to think of a sequential run as a run with 0 jumps, we measure a jump over $k$ letters as distance $k-1$ (either to the left or to the right). This is captured as follows.

For $k\in \bbZ$, define $\sem{k}=|k|-1$. Consider a word $w\in \Sigma^*$ with $|w|=n$, and let $\vec{a}=(a_0, a_1, a_2, \ldots, a_n, a_{n+1})$ be a jump sequence, then we lift the notation above and write $\sem{\vec{a}}=\sum_{i=1}^{n+1}\sem{a_i-a_{i-1}}$.

\begin{defi}[Absolute-Distance Semantics]
\label{def:absolute distance}
For a word $w\in \Sigma^*$ with $|w|=n$ we define
\[
\cA_{\Abs}(w)=
\min\{\sem{\vec{a}}\mid \vec{a} \text{ is a jump sequence, and } w_{\vec{a}}\in \regL(\cA)\}
\]
(recall that $\min \emptyset=\infty$ by definition).
\end{defi}

\paragraph*{The Reversal Semantics}
In the reversal semantics, the cost of a run is the number of times the head changes direction in the corresponding jump sequence.
Consider a word $w\in \Sigma^*$ with $|w|=n$, and let $\vec{a}=(a_0, a_1, a_2, \ldots, a_n, a_{n+1})$ be a jump sequence, we define 
\[\numrev(\vec{a})=|\{i\in \{1,\ldots,n\}\mid (a_i>a_{i-1}\wedge a_i>a_{i+1}) \vee (a_i<a_{i-1}\wedge a_i<a_{i+1})\}|\]

 
\begin{defi}[Reversal Semantics]
\label{def:reversal semantics}
For a word $w\in \Sigma^*$ with $|w|=n$ we define
\[
\cA_{\Rev}(w)=
\min\{\numrev(\vec{a})\mid \vec{a} \text{ is a jump sequence, and } w_{\vec{a}}\in \regL(\cA)\}\]
\end{defi}

\paragraph*{The Hamming Semantics}
In the Hamming measure, the cost of a word is the minimal number of coordinates of $w$ that need to be changed in order for the obtained word to be accepted by $\cA$ (sequentially, as an \NFA), so that the changed word is a permutation of $w$.

Consider two words $x,y\in \Sigma^*$ with $|x|=|y|=n$ such that $x\sim y$, we define the \emph{Hamming Distance} between $x$ and $y$ as $d_H(x,y)=|\{i\mid x_i\neq y_i\}|$. 

\begin{defi}[Hamming Semantics]
\label{def:hamming semantics}
 For a word $w\in \Sigma^*$ we define
 \[
\cA_{\Ham}(w)=\min\{d_H(w',w)\mid w'\in L(\cA), w'\sim w\}
\]
\end{defi}

\paragraph*{The Maximum Jump Semantics}
In the maximum jump semantics, the cost of a run is the size of the maximum jump made by the head. Let $\vec{a}=(a_0, a_1, a_2, \ldots, a_n, a_{n+1})$ be a jump sequence. We define $\sem{\vec{a}}_\Max=\max_{i=1}^{n+1}\sem{a_i-a_{i-1}}$.

\begin{defi}[Maximum Jump Semantics]
\label{def:maxjump semantics}
For a word $w\in \Sigma^*$ with $|w|=n$ we define
\[
\cA_{\Max}(w)=
\min\{\sem{\vec{a}}_\Max\mid \vec{a} \text{ is a jump sequence, and } w_{\vec{a}}\in \regL(\cA)\}
\]
\end{defi}

\begin{rem}
    \label{rmk:semantics semantics}
    Note that the definitions of all the semantics are independent of the $\NFA$, and only refer to its language. We can therefore refer to the cost of a word in a language according to each semantics, rather than the cost of a word in a concrete automaton. 
\end{rem}

\subsection{Quantitative Decision Problems}
In the remainder of the paper we focus on quantitative variants of the standard Boolean decision problems pertaining to the jumping semantics. Specifically, we consider the following problems for each semantics $\Sem\in \{\Abs,\Ham,\Rev,\Max\}$.
\begin{itemize}
    \item \probMembership: Given a \JFA $\cA$, $k\in \bbN$ and a word $w$, decide whether $\cA_{\Sem}(w)\le k$.
    \item \probDomkBoundedness (for a fixed $k$): Given a \JFA $\cA$, decide whether $\forall w\in \jL(\cA)\  \cA_{\Sem}(w)\le k$.
    \item \probDomParamBoundedness: Given a \JFA $\cA$ and $k\in \bbN$, decide whether $\forall w\in \jL(\cA)\  \cA_{\Sem}(w)\le k$.
\end{itemize}
We also pay special attention to the setting where $\jL(\cA)=\Sigma^*$, in which case we refer to these problems as \probkBoundedness and \probParamBoundedness.
For example,
    in \probParamBoundedness we are given a \JFA $\cA$ and $k\in \bbN$ and the problem is to decide whether $\cA_{\Sem}(w)\le k$ for all words $w\in \Sigma^*$.

The boundedness problems can be thought of as quantitative variants of Boolean universality (i.e., is the language equal to $\Sigma^*$). Observe that the problems above are not fully specified, as the encoding of $k$ (binary or unary) when it is part of the input may effect the complexity. We remark on this when it is relevant. 
Note that the emptiness problem is absent from the list above. Indeed, a natural quantitative variant would be: is there a word $w$ such that $\cA_{\Sem}(w)\le k$. This, however, is identical to Boolean emptiness, since $\regL(\cA)\neq \emptyset$ if and only if there exists $w$ such that $\cA_{\Sem}(w)=0$. We therefore do not consider this problem.
Another problem to consider is boundedness when $k$ is existentially quantified. We elaborate on this problem in~\cref{sec:discussion}.

\section{The Absolute-Distance Semantics}
\label{sec:absolute distance}
The first semantics we investigate is $\Abs$, and we start by showing that (the decision version of) computing its value for a given word is \cxNPc. This is based on bounding the distance with which a word can be accepted.
\begin{lem}
\label{lem:dist max val}
Consider a JFA $\cA$ and $w\in \jL(\cA)$ with $|w|=n$, then $\cA_{\Abs}(w)\le  n^2$.
\end{lem}
\begin{proof}
For $n=0$, i.e., $w=\epsilon$, the lemma clearly holds with $\cA_{\Abs}(\epsilon)=0$. Assume $n>0$, we actually show a strict inequality.
Since $w\in \jL(\cA)$, there exists a jump sequence $\vec{a}=(a_0, a_1, a_2, \ldots, a_n, a_{n+1})$ such that $w_{\vec{a}}\in \regL(\cA)$. Therefore, $\cA_\Abs(w)\le \sem{\vec{a}}$. Observe that $|a_i-a_{i-1}|\le {n}$ for all $i\in \{1,\ldots,n+1\}$, since there is no jump from $0$ to $n+1$ (since $a_0=0$ and $a_n=n+1$). The following concludes the proof:
\[
\sem{\vec{a}}=\sum_{i=1}^{n+1}\sem{a_i-a_{i-1}}=\sum_{i=1}^{n+1}|a_i-a_{i-1}|-1\le \sum_{i=1}^{n+1}n-1=(n+1)(n-1)<n^2
\qedhere\]
\end{proof}
We can now prove the complexity bound for computing the absolute distance, as follows.
\begin{thm}[Absolute-Distance \probMembership is $\cxNPc$]
    \label{thm:dist membership NPcomplete}
    The problem of deciding, given $\cA, w$ and $k\in \bbN$, whether $\cA_\Abs(w)\le k$, is \cxNPc.
\end{thm}
\begin{proof}
    In order to establish membership in \cxNP, note that by~\cref{lem:dist max val}, we can assume $k\le n^2$, as otherwise we can set $k=n^2$. Then, it is sufficient to nondeterministically guess a jump sequence $\vec{a}$ and to check that $w_{a_1}\cdots w_{a_n}\in \regL(\cA)$ and that $\sem{\vec{a}}\le k$. Both conditions are easily checked in polynomial time, since $k$ is polynomially bounded.

    Hardness in \cxNP follows by reduction from (Boolean) membership in \JFA: it is shown in~\cite{fernau2017characterization} that deciding whether $w\in \jL(\cA)$ is \cxNPh. We reduce this problem by outputting, given $\cA$ and $w$, the same $\cA$ and $w$ with the bound $k=n^2$. The reduction is correct by~\cref{lem:dist max val} and the fact that if $w\notin \jL(\cA)$ then $\cA_\Abs(w)=\infty$.
\end{proof}

\subsection{Decidability of Boundedness Problems for ABS}
\label{sec:abs decidability}
We now turn our attention to the boundedness problems. 
Consider a \JFA $\cA$ and $k\in \bbN$. 
Intuitively, our approach is to construct an \NFA $\cB$ that simulates, while reading a word $w\in \Sigma^*$, every jump sequence of $\cA$ on $w$ whose absolute distance is at most $k$. The crux of the proof is to show that we can indeed bound the size of $\cB$ as a function of $k$. 
At a glance, the main idea here is to claim that since the absolute distance is bounded by $k$, then $\cA$ cannot make large jumps, nor many small jumps. Then, if we track a sequential head going from left to right, then the jumping head must always be within a bounded distance from it. We now turn to the formal arguments. Fix a \JFA $\cA=\tup{\Sigma,Q,\delta,Q_0,\alpha}$.

To understand the next lemma, imagine $\cA$'s jumping head while taking the $j^{th}$ step in a run on $w$ according to a jump sequence $\vec{a}=(a_0, a_1, a_2, \ldots, a_n, a_{n+1})$. Thus, the jumping head points to the letter at index $a_j$. Concurrently, imagine a ``sequential'' head (reading from left to right), which points to the $j^{th}$ letter in $w$. Note that these two heads start and finish reading the word at the same indices $a_0=0$ and $a_{n+1} = n+1$. It stands to reason that if at any step while reading $w$ the distance between these two heads is large, the cost of reading $w$ according to $\vec{a}$ would also be large, as there would need to be jumps that bridge the gaps between the heads. 
The following lemma formalizes this idea.

\begin{lem} 
\label{lem:abs window}
 Consider a jump sequence $\vec{a}=(a_0, a_1, a_2, \ldots, a_n, a_{n+1})$. For every $1\le j\le n$ it holds that $\sem{\vec{a}}\ge |a_j-j|$.
 \end{lem}

\begin{proof}
\newcommand{\xdownarrow}[1]{%
  {\left\downarrow\vbox to #1{}\right.\kern-\nulldelimiterspace}
}
\newcommand{\arrowpointer}[2]{\underset{\mathclap{\substack{\color{blue} \big \downarrow \\ \textcolor{blue}{#1}}}}{#2}}
Let $1\le j\le n$. First, assume that $a_j \ge j$ and consider the sum $\sum_{i=1} ^j \sem{a_i - a_{i-1}} \le \sem{\vec{a}}$. From the definition of $\sem{\cdot}$ we have $\sum_{i=1} ^j \sem{a_i - a_{i-1}} = \left(\sum_{i=1} ^j |a_i - a_{i-1}|\right) - j$,
and we conclude that in this case $\sem{\vec{a}}\ge |a_j-j|$ by the following:
\[\left(\sum_{i=1} ^j |a_i - a_{i-1}|\right) -j \arrowpointer{\text{triangle inequality}}{\ge} \left|\sum_{i=1} ^j a_i - a_{i-1}\right|-j \arrowpointer{\text{telescopic sum}}{=} |a_j-a_0| - j \arrowpointer{a_0=0}{=} a_j - j\arrowpointer{a_j\ge j}{=}|a_j-j|\]
The direction $a_j<j$ is proved by looking at the sum of the \emph{last} $j$ elements:
assume $a_j < j$, and consider the sum $\sum_{i=j+1}^{n+1} \sem{a_i - a_{i-1}} \le \sem{\vec{a}}$. From the definition of $\sem{\cdot}$ we have 
 \[\sum_{i=j+1} ^{n+1} \sem{a_i - a_{i-1}} = \left(\sum_{i=j+1} ^{n+1} |a_i - a_{i-1}|\right) - (n+1 - (j+1) + 1)=\left(\sum_{i=j+1} ^{n+1} |a_i - a_{i-1}|\right) - (n + 1 - j)\]
 Similarly to the previous case, from the triangle inequality we have
 \[
 \begin{split}
 &\left(\sum_{i=j+1} ^{n+1} |a_i - a_{i-1}|\right) - (n+1 - j) \ge |a_{n+1}-a_j| - (n+1- j) =\\
 &n+1 - a_j - (n+1-j)=j-a_j=|a_j-j|
 \end{split}
 \] 
 where we use the fact that $a_{n+1}=n+1>a_j$, and our assumption that $a_j<j$. This again concludes that $\sem{\vec{a}}\ge |a_j-j|$.
\end{proof}
From~\cref{lem:abs window} we get that in order for a word $w$ to attain a small cost, it must be accepted with a jumping sequence that stays close to the sequential head. More precisely:
\begin{cor}
    \label{cor:dist max cost}
    Let $k\in \bbN$ and consider a word $w$ such that $\cA_{\Abs}(w)\le k$, then there exists a jumping sequence $\vec{a}=(a_0, a_1, a_2, \ldots, a_n, a_{n+1})$ such that $w_{\vec{a}}\in \regL(\cA)$ and for all $1\le j\le n$ we have $|a_j-j|\le k$.
\end{cor}

We now turn to the construction of an \NFA that recognizes the words whose cost is at most $k$. 
\begin{lem}
    \label{lem:dist NFA construction}
    Let $k\in \bbN$. We can effectively construct an \NFA $\cB$ such that $\regL(\cB)\!=\!\{w\in \Sigma^* | \cA_{\Abs}(w)\le k\}$.
\end{lem}
\begin{proof}
    Let $k\in \bbN$. Intuitively, $\cB$ works as follows: it remembers in its states a window of size $2k+1$ centered around the current letter (recall that as an \NFA, it reads its input sequentially). The window is constructed by nondeterministically guessing (and then verifying) the next $k$ letters, and remembering the last $k$ letters. 

$\cB$ then nondeterministically simulates a jumping sequence of $\cA$ on the given word, with the property that the jumping head stays within distance $k$ from the sequential head. This is done by marking for each letter in the window whether it has already been read in the jumping sequence, and nondeterministically guessing the next letter to read, while keeping track of the current jumping head location, as well as the total cost incurred so far.
After reading a letter, the window is shifted by one to the right. 
If at any point the window is shifted so that a letter that has not been read by the jumping head shifts out of the $2k+1$ scope, the run rejects. Similarly, if the word ends but the guessed run tried to read a letter beyond the length of the word, the run rejects.
The correctness of the construction follows from~\cref{cor:dist max cost}.
We now turn to the formal details.
Recall that $\cA=\tup{\Sigma,Q,\delta,Q_0,\alpha}$. We define $\cB=\tup{\Sigma,Q',\delta',Q_0',\beta}$ as follows.

The state space of $\cB$ is 
\[Q'=Q\times (\Sigma\times \{?,\checkmark\})^{-k,\ldots,k}\times\{-k,\ldots,k\}\times \{0,\ldots,k\}\] 
We denote a state of $\cB$ as $(q,f,j,c)$ where $q\in Q$ is a state of $\cA$, $f:\{-k,\ldots,k\}\to \Sigma\times\{?,\checkmark\}$ represents a window of size $2k+1$ around the sequential head, where $\checkmark$ marks letters that have already been read by $\cA$ (and $?$ marks the others), $j$ represents the index of the head of $\cA$ relative to the sequential head, and $c$ represents the cost incurred thus far in the run. We refer to the components of $f$ as $f(j)=(f(j)_1,f(j)_2)$ with $f(j)_1\in \Sigma$ and $f(j)_2\in \{?,\checkmark\}$.

The initial states of $\cB$ are 
\[Q'_0=\left\{(q,f,j,j-1)\mid q\in Q_0\wedge j>0 \wedge (f(i)_2=\checkmark \iff i\le 0)\right\}\] 
That is, all states where the state of $\cA$ is initial, the location of the jumping head is some $j>0$ incurring a cost of $j-1$ (i.e., the initial jump $\cA$ makes), and the window is guessed so that everything left of the first letter is marked as already-read (to simulate the fact that $\cA$ cannot jump to the left of the first letter).

The transitions of $\cB$ are defined as follows. Consider a state $(q,f,j,c)$ and a letter $\sigma\in \Sigma$, then $(q',f',j',c')\in \delta'((q,f,j,c),\sigma)$ if and only if the following hold (see~\cref{fig:abs transition} for an illustration):
\begin{itemize}
    \item $f(1)_1=\sigma$. That is, we verify that the next letter in the guessed window is indeed correct.
    \item $f(-k)_2=\checkmark$. That is, the leftmost letter has been read. Otherwise by~\cref{cor:dist max cost} the cost of continuing the run must be greater than $k$.
    \item $f(j)_2\neq \checkmark$ and $f'(j-1)=\checkmark$ (if $j>-k$). That is, the current letter has not been previously read, and will be read from now on (note that index $j$ before the transition corresponds to index $j-1$ after).
    \item $q'=\delta(q,f(j)_1)$, i.e. the state of $\cA$ is updated according to the current letter.
    \item $c'=c+|j'+1-j|-1$, since $j'$ represents the index in the shifted window, so in the ``pre-shifted'' tape this is actually index $j+1$. We demonstrate this in~\cref{fig:abs transition}. Also, $c'\le k$ by the definition of $Q$.
    \item $f'(i)=f(i+1)$ for $i<k$. That is, the window is shifted and the index $f'(k)$ is nondeterministically guessed\footnote{The guess could potentially be $\checkmark$, but this is clearly useless.}.
\end{itemize}

\newcommand*{\mybox}[1]{%
  \framebox{\raisebox{0pt}[0.5\baselineskip][0\baselineskip]{%
    \hbox to 10pt{\hss#1\hss}}}}

\begin{figure}[ht]
    \centering
    \begin{tikzpicture}[auto,node distance=2.8cm,scale=1]
        \node (main1) [draw=none] at (0,0) {\mybox{$\alpha\checkmark$}\mybox{$\beta?$}\mybox{$\gamma\checkmark$}\mybox{$\epsilon?$}\mybox{$\mu\checkmark$}\mybox{$\eta?$}\mybox{$\xi?$}\mybox{$\zeta\checkmark$}\mybox{$\theta?$}};
        \foreach \i in {-4,...,4}
        {
            \pgfmathtruncatemacro{\x}{\i*17};
            \node (let\i) at (\x pt,12pt) {\footnotesize \color{gray} $\i$};
        }
        \draw[->,gray,bend left] (let-3) to [bend left] (let2);
        \node[inner sep=7pt] (seq) at (0,0) {};
        \node[inner sep=7pt] (jump) at (-50pt,0) {};
        \draw[->,dashed] (0,-20pt) -- (seq);
        \draw[->] (-50pt,-20pt) -- (jump);
        \node (impl) at (3.7,0) {\huge $\implies$};
    \end{tikzpicture}
    \begin{tikzpicture}[auto,node distance=2.8cm,scale=1]
        \node (main1) [draw=none] at (0,0) {\mybox{$\beta\checkmark$}\mybox{$\gamma\checkmark$}\mybox{$\epsilon?$}\mybox{$\mu\checkmark$}\mybox{$\eta?$}\mybox{$\xi?$}\mybox{$\zeta\checkmark$}\mybox{$\theta?$}\mybox{$\psi?$}};
        \foreach \i in {-4,...,4}
        {
            \pgfmathtruncatemacro{\x}{\i*17};
            \node (let\i) at (\x pt,12pt) {\footnotesize \color{gray} $\i$};
        }
        \node[inner sep=7pt] (seq) at (0,0) {};
        \node[inner sep=7pt] (jump) at (18pt,0) {};
        \draw[->,dashed] (0,-20pt) -- (seq);
        \draw[->] (18pt,-20pt) -- (jump);
    \end{tikzpicture}
    \caption{A single transition in the construction of~\cref{lem:dist NFA construction}. The dashed arrow signifies the sequential head, the full arrow is the ``imaginary'' jumping head. Here, the head jumps from $-3$ to $2$, incurring a cost of $4$, but in the indexing after the transition $\xi$ is at index $1$, thus the expression given for $c'$ in the construction. Note that the letter being read must be $\mu$, and that $\alpha$ must be checked, otherwise the run has failed.}
    \label{fig:abs transition}
\end{figure}
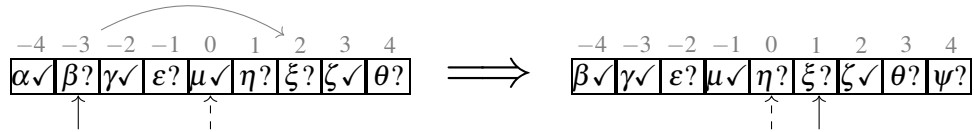
\noindent 
Finally, the accepting states of $\cB$ are $\beta=\{(q,f,1,c)\mid q\in \alpha\wedge f(j)_2=? \text{ for all }j>0\}$. That is, the state of $\cA$ is accepting, the overall cost is at most $k$, the location of the jumping head matches the sequential head (intuitively, location $n+1$), and no letter beyond the end of the tape has been used.
    
It is easy to verify that $\cB$ indeed guesses a jump sequence and a corresponding run of $\cA$ on the given word, provided that the jumping head stays within distance $k$ of the sequential head. By~\cref{cor:dist max cost}, this restriction is complete, in the sense that if $\cA_{\Abs}(w)\le k$ then there is a suitable jump sequence under this restriction with which $w$ is accepted.
\end{proof}

We can now readily conclude the decidability of the boundedness problems for 
the $\Abs$ semantics. The proof 
makes use of the decidability of emptiness for Parikh Automata~\cite{klaedtke2003monadic}.
\begin{thm}
\label{thm:abs decidable boundedness}
The following problems are decidable for the $\Abs$ semantics: \probDomkBoundedness, \probDomParamBoundedness, \probkBoundedness and \probParamBoundedness.
\end{thm}
 \begin{proof}
    Consider a \JFA $\cA$ and $k\in \bbN$ ($k$ is either fixed or given as input, which does not affect decidability), and let $\cB$ be the \NFA constructed as per~\cref{lem:dist NFA construction}. 
    In order to decide \probkBoundedness and \probParamBoundedness, observe that $\cA_\Abs(w)\le k$ for every word $w\in \Sigma^*$  if and only if $\regL(\cB)=\Sigma^*$. Since the latter is decidable for \NFA, we have decidability.

    Similarly, in order to decide \probDomkBoundedness and \probParamBoundedness, observe that $\cA_\Abs(w)\le k$ for every word $w\in \jL(\cA)$ if and only if $\jL(\cA)\subseteq \regL(\cB)$. We can decide whether the latter holds by constructing the PA $(\overline{\cB},\cA)$ where $\overline{\cB}$ is an \NFA for the complement of $\regL(\cB)$, and checking emptiness. Since emptiness for PA is decidable~\cite{klaedtke2003monadic}, we conclude decidability.
\end{proof}
With further scrutiny, we see that the size of $\cB$ constructed as per~\cref{lem:dist NFA construction} is polynomial in the size of $\cA$ and single-exponential in $k$. Thus, \probkBoundedness is in fact decidable in $\cxPSPACE$, whereas \probParamBoundedness is in $\cxEXPSPACE$ and $\cxEXPEXPSPACE$ for $k$ given in unary and binary, respectively. For the non-universal problems we do not supply upper complexity bounds, as these depend on the decidability for PA containment, for which we only derive decidability from~\cite{klaedtke2003monadic}.

\subsection{\textsf{PSPACE-Hardness} of Boundedness for ABS}
\label{sec:abs hardness}
In the following, we complement the decidability result of~\cref{thm:abs decidable boundedness} by showing that already \probkBoundedness is $\cxPSPACEh$, for every $k\in \bbN$.

We first observe that the absolute distance of every word is even. In fact, this is true for every jumping sequence.
\begin{lem}
\label{lem:dist is even}
Consider a jumping sequence $\vec{a}=(a_0, a_1, \ldots, a_n, a_{n+1})$, then $\sem{\vec{a}}$ is even.
\end{lem}
\begin{proof}
Observe that the parity of $|a_{i}-a_{i-1}|$ is the same as that of $a_i-a_{i-1}$. It follows that the parity of $\sem{\vec{a}}=\sum_{i=1}^{n+1}\sem{a_i-a_{i-1}}=\sum_{i=1}^{n+1}|a_i-a_{i-1}|-1$ is the same as that of 
\[\sum_{i=1}^{n+1} (a_i-a_{i-1}-1)=\left(\sum_{i=1}^{n+1} a_i-a_{i-1}\right)-(n+1)=n+1-(n+1)=0\]
and is therefore even (the penultimate equality is due to the telescopic sum).
\end{proof}

We say that $\cA_\Abs$ is \emph{$k$-bounded} if $\cA_\Abs(w)\le k$ for all $w\in \Sigma^*$.
We are now ready to prove the hardness of \probkBoundedness. Observe that for a word $w\in \Sigma^*$ we have that $\cA_\Abs(w)=0$ if and only if $w\in \regL(\cA)$ (indeed, a cost of $0$ implies that an accepting jump sequence is the sequential run $0,1,\ldots,|w|+1$). In particular, we have that $\cA_\Abs$ is $0$-bounded if and only if $\regL(\cA)=\Sigma^*$. Since the universality problem for \NFAs is \cxPSPACEc, this readily proves that \probUnivBoundedness{$0$} is \cxPSPACEh. 
Note, however, that this does \emph{not} imply that \probkBoundedness is also $\cxPSPACEh$ for other values of $k$, and that the same argument fails for $k>0$. We therefore need a slightly more elaborate reduction.

\newcommand{\fs}{\heartsuit}
\begin{lem}
\label{lem:dist k boundedness PSPACE hard}
    For $\Abs$ the \probkBoundedness and \probDomkBoundedness problems are \cxPSPACEh for every $k\in \bbN$.
\end{lem}
\begin{proof}
We begin with an intuitive sketch the proof for \probkBoundedness. The case of \probDomkBoundedness requires slightly more effort. 
By~\cref{lem:dist is even}, we can assume without loss of generality that $k$ is even. Indeed, if there exists $m\in \bbN$ such that $\cA_\Abs(w)\le 2m+1$ for every $w\in \Sigma^*$, then by~\cref{lem:dist is even} we also have $\cA_\Abs(w)\le 2m$. Therefore, we assume $k=2m$ for some $m\in \bbN$.

We reduce the universality problem for $\NFAs$ to the  \probUnivBoundedness{$2m$} problem. Consider an \NFA $\cA=\tup{Q,\Sigma,\delta,Q_0,\alpha}$, and let $\fs\notin \Sigma$ be a fresh symbol. 
Intuitively, we obtain from $\cA$ an \NFA $\cB$ over the alphabet $\Sigma\cup \{\fs\}$ such that $w\in \regL(\cB)$ if and only if the following hold:
\begin{enumerate}
    \item Either $w$ does not contain exactly $m$ occurrences of $\fs$, or
    \item $w$ contains exactly $m$ occurrences of $\fs$, but does not start with $\fs$, and $w|_\Sigma\in \regL(\cA)$ (where $w|_\Sigma$ is obtained from $w$ by removing all occurrences of $\fs$).
\end{enumerate}
We then have the following: if $\regL(\cA)=\Sigma^*$, then for every $w\in (\Sigma\cup \{\fs\})^*$ if $w\in \regL(\cB)$ then $\cB_\Abs(w)=0\le 2m$, and if $w\notin \regL(\cB)$ then $w$ starts with $\fs$ but has exactly $m$ occurrences of $\fs$. Thus, jumping to the first occurrence of a letter in $\Sigma$ incurs a cost of at most $m$, and reading the skipped $\fs$ symbols raises the cost to at most $2m$. From there, $w$ can be read consecutively and be accepted since $w|_\Sigma\in \regL(\cA)$. So again $\cB_\Abs(w)\le 2m$, and $\cB$ is $2m$-bounded. 

Conversely, if $\regL(\cA)\neq \Sigma^*$, take $x\notin \regL(\cA)$ such that $x\neq \epsilon$ (we show in the details below why this can be assumed), and consider the word $w=\fs^m x$. We then have $w\notin \regL(\cB)$, and moreover -- in order to accept $w$ (if at all possible), $\cB$ first needs to jump over the initial $\fs^m$, guaranteeing a cost of at least $2m$ ($m$ for the jump and another $m$ to later read the $\fs^m$ prefix), and needs at least one more jump to accept $x$, since $x\notin \regL(\cA)$. Thus, $\cB_\Abs(w)>2m$, so $\cB$ is not $2m$-bounded.
%
%
We now turn to the precise details of the reduction, differing from the intuition above mainly by the pesky presence of $\epsilon$. Given $\cA$, we first check whether $\epsilon\in \regL(\cA)$ (i.e., we check whether $Q_0\cap \alpha \neq \emptyset$). If $\epsilon\notin \regL(\cA)$, then $\regL(\cA)\neq \Sigma^*$ and we output some fixed unbounded automaton $\cB$ (e.g., as in~\cref{xmp:semantics ab simple,xmp:UUUU}).
Otherwise, we proceed to construct $\cB$ such that $w\in \regL(\cB)$ if and only if either Conditions 1 or 2 above hold, or:
\begin{enumerate}
\setcounter{enumi}{2}
    \item $w=\fs^m$ (think of this as an exception to Condition 2 where $w|_\Sigma=\epsilon$).
\end{enumerate}
Constructing $\cB$ from $\cA$ is straightforward by taking roughly $m$ copies of $\cA$ to track the number of $\fs$ in the word. In particular, the reduction is in polynomial time.

We claim that $\regL(\cA)=\Sigma^*$ if and only if $\cB$ is $2m$-bounded.
For the first direction, assume $\regL(\cA)=\Sigma^*$. As mentioned above, for every $w\in (\Sigma\cup \{\fs\})^*$ if $w\in \regL(\cB)$ then $\cB_\Abs(w)=0\le 2m$. If $w\notin \regL(\cB)$ then $w$ starts with $\fs$ but has exactly $m$ occurrences of $\fs$, and in addition, $w\neq \fs^m$ so there is some $m'\le m+1$ such that the $m'$-th letter of $w$ is in $\Sigma$. Then, the following jump sequence $\vec{a}$ causes $\cB$ to accept $w_{\vec{a}}$: 
\[\vec{a}=(0,m',1,2,\ldots,m'-1,m'+1,m'+2,\ldots,|w|,|w|+1)\]
Indeed, $w_{\vec{a}}$ does not start with $\fs$, has exactly $m$ occurrences of $\fs$, and $w|_{\Sigma}\in \regL(\cA)=\Sigma^*$, so Condition 2 holds. Finally, note that $\sem{\vec{a}}=m'-1+m'-2+1=2(m'-1)\le 2m$ (since the only non-zero jumps are $0$ to $m'$, $m'$ to $1$, and $m'-1$ to $m'+1$).

For the converse, assume $\regL(\cA)\neq \Sigma^*$. If $\epsilon\notin \regL(\cA)$, then correctness follows by the treatment of this case above. Otherwise, let $x\notin \regL(\cA)$ with $x\neq \epsilon$, and consider $w=\fs^m x$. We claim that $\cB_\Abs(w)>2m$. Indeed, $w\notin \regL(\cB)$, and therefore $\cB_\Abs(w)>0$. If $\cB_\Abs(w)=\infty$ then we are done. Otherwise, let $\vec{a}$ be a jump sequence such that $w_\vec{a}\in \regL(\cB)$. Then $a_1\ge m+1$, contributing a cost of at least $m$ to $\vec{a}$. It can be easily seen by induction over $m$ that in order for $\vec{a}$ to cover the entries $1,\ldots, m$ starting at position $m+1$ and ending at position $m+2$, it requires cost of at least another $m$. Then, however, in order for $w_\vec{a}$ to be accepted by $\cB$, it must hold that $w_\vec{a}|_\Sigma\neq x$, so $\vec{a}$ is not the identity starting from $m+1$. It therefore has an additional cost of at least $1$. Thus, $\sem{\vec{a}}>2m$. In particular, $\cB_{\Abs}(w)>2m$, so $\cB$ is not $2m$-bounded, and we are done.
\end{proof}
\cref{lem:dist k boundedness PSPACE hard} shows hardness for fixed $k$, and in particular when $k$ is part of the input. Thus, \probParamBoundedness and \probDomParamBoundedness are also $\cxPSPACEh$, and $\probkBoundedness$ is $\cxPSPACEc$. Also,  \probParamBoundedness is in $\cxEXPSPACE$ and $\cxEXPEXPSPACE$ for $k$ given in unary and binary, respectively. 
%
 

\section{The Reversal Semantics}
\label{sec:reversal}
We now study the reversal semantics. Recall from~\cref{def:reversal semantics} that for a \JFA $\cA$ and a word $w$, the cost $\cA_\Rev(w)$  is the minimal number of times the jumping head changes ``direction'' in a jump sequence for which $w$ is accepted. 

Consider a word $w$ with $|w|=n$ and a jump sequence $\vec{a}=(a_0, a_1, a_2, \ldots, a_n, a_{n+1})$. We say that an index $1\le i\le n$ is a \emph{turning index} if $a_i>a_{i-1}$ and $a_i>a_{i+1}$ (i.e., a right-to-left turn) or if $a_i<a_{i-1}$ and $a_i<a_{i+1}$ (i.e., a left-to-right turn). We denote by $\turnindices(\vec{a})$ the set of turning indices of $\vec{a}$. 

For example, consider the jump sequence $(\overset{a_0}{0},\overset{a_1}{2},\overset{a_2}{3},\overset{a_3}{5},\overset{a_4}{7},\overset{a_5}{4},\overset{a_6}{1},\overset{a_7}{6},\overset{a_8}{8})$, then $\turnindices(\vec{a})=\{4,6\}$.
Note that the cost of $w$ is then $\cA_\Rev(w)=\min\{|\turnindices(\vec{a})| \mid w_{\vec{a}}\in \regL(\cA)\}$.
Viewed in this manner, we have that $\cA_\Rev(w)\le |w|$, and computing $\turnindices(\vec{a})$ can be done in polynomial time. Thus, analogously to~\cref{thm:dist membership NPcomplete} we have the following.
\begin{thm}[Reversal \probMembership is $\cxNPc$]
    \label{thm:rev membership NPc}
    The problem of deciding, given $\cA$ and $k$, whether $\cA_\Rev(w)\le k$ is $\cxNPc$.
\end{thm}
\begin{rem}
    \label{rmk:rev even}
    For every jump sequence $\vec{a}$ we have that $|\turnindices(\vec{a})|$ is even, since the head starts at position $0$ and ends at $n+1$, where after an odd number of turning points the direction is right-to-left, and hence cannot reach $n+1$.
\end{rem}

\subsection{Decidability of Boundedness Problems for REV}
\label{sec:rev boundedness decidable}

We begin by characterizing the words accepted using at most $k$ reversals as a shuffle of subwords and reversed-subwords, as follows.
Let $x,y\in \Sigma^*$ be words, we define their \emph{shuffle} to be the set of words obtained by interleaving parts of $x$ and parts of $y$. Formally:
\[
x\shuffle y=\{s_1\cdot t_1\cdot s_2\cdot t_2\cdots s_k\cdot t_k\mid \forall i\ s_i,t_i\in \Sigma^*\wedge  x=s_1\cdots s_k\wedge y=t_1\cdots t_k\}
\]
For example, if $\color{red} x=aab$ and $\color{blue} y=cd$ then ${\color{red} x}\shuffle {\color{blue} y}$ contains the words ${\color{red}aab}{\color{blue}cd}$, ${\color{red}a}{\color{blue}c}{\color{red}ab}{\color{blue}d}$, ${\color{blue}c}{\color{red}aa}{\color{blue}d}{\color{red}b}$, among others (the colors reflect which word each subword originated from).
Note that the subwords may be empty, e.g., ${\color{blue}c}{\color{red}aa}{\color{blue}d}{\color{red}b}$ can be seen as starting with $\color{red}\epsilon$ as a subword of $\color{red} x$. It is easy to see that $\shuffle$ is an associative operation, so it can be extended to any finite number of words. 

The following lemma states that, intuitively, if $\cA_\Rev(w)\le k$, then $w$ can be decomposed to a shuffle of at most $k+1$ subwords of itself, where all the even ones are reversed (representing the left-reading subwords). 
\begin{lem} 
\label{lem: turning subwords shuffle}
Let $k\in \bbN$. Consider an \NFA $\cA$ and a word $w\in \Sigma^*$. Then $\cA_\Rev(w)\le k$ if and only if there exist words $s_1, s_2,\ldots, s_{k+1}\in \Sigma^*$ such that the following hold.
\begin{enumerate}
    \item $s_1 s_2 \ldots s_{k+1}\in \regL(\cA)$.
    \item $w\in s_1 \shuffle {s_2}^R \shuffle s_3 \shuffle {s_4}^R \shuffle \ldots \shuffle s_{k+1}$ (where $s_i^R$ is the reverse of $s_i$).
\end{enumerate}
\end{lem}
\begin{proof}
For the first direction, assume $\cA_\Rev(w)\le k$, so there exists a jump vector $\vec{a}$ such that $|\turnindices(\vec{a})| \le k$ and $w_\vec{a}\in \regL(\cA)$. 
Write $\turnindices(\vec{a})=\{i_1,i_2,\ldots,i_\ell\}$ where $i_1<i_2<\ldots<i_\ell$, and set $i_0=0$ and $i_{\ell+1}=n+1$. Then, for every $1\le j\le \ell+1$, the \emph{$j$-th turning subword} is $s_j=w_{a_{i_{j-1}}}w_{a_{i_{j-1}+1}}\cdots w_{a_{i_{j}}}$ (for $j=\ell+1$ we end with $w_{a_{i_{\ell+1}-1}}$). If $\ell+1<k+1$ we define the remaining subwords $s_{\ell+2},\ldots,s_{k+1}$ to be $\epsilon$. To avoid cumbersome indexing, we assume $\ell+1=k+1$ in the following.

It is now easy to see that Conditions 1 and 2 hold for $s_1,\ldots,s_{k+1}$. 
Indeed, by definition we have $s_1\cdots s_{k+1}\in \regL(\cA)$, so Condition 1 holds. 
For Condition 2, observe that for every $1\le i\le k+1$, if $i$ is odd, then $s_i$ consists of an ascending sequence of letters, and if $i$ is even then $s_i$ is a descending sequence. Since the $s_i$ form a partition of the letters of $w$, we can conclude that  $w\in s_1 \shuffle {s_2}^R \shuffle s_3 \shuffle {s_4}^R \shuffle \ldots \shuffle s_{k+1}$ (by shuffling the letters of these words to form exactly the sequence of indices $1,\ldots, |w|$).

For the converse, consider words $s_1,\ldots,s_{k+1}$ such that Conditions 1 and 2 hold.
By Condition 2, we see that the word $s_1 s_2\ldots s_{k+1}$ is a permutation of $w$, and moreover -- from the way $w$ is obtained in $s_1 \shuffle {s_2}^R \shuffle s_3 \shuffle {s_4}^R \shuffle \ldots \shuffle s_{k+1}$ we can extract a jump sequence $\vec{a}$ such that $w_{\vec{a}}=s_1,\ldots,s_{k+1}$ and such that the turning subwords of $\vec{a}$ are exactly $s_1,s_2^R,\ldots,s_{k+1}$. Indeed, this follows from the same observation as above -- for odd $i$ we have that $s_i$ is an increasing sequence of indices, and for even $i$ it is decreasing. 
In particular, $|\turnindices(\vec{a})|\le k$, so $\cA_{\Rev}(w)\le k$.
\end{proof}
Using the characterization in~\cref{lem: turning subwords shuffle}, we can now construct a corresponding $\NFA$, by intuitively guessing the shuffle decomposition and running copies of $\cA$ and its reverse in parallel. 
\begin{lem}
\label{lem:rev k construction}
    Let $k\in \bbN$ and consider a \JFA $\cA$. We can effectively construct an \NFA $\cB$ such that $\regL(\cB)=\{w\in \Sigma^*\mid \cA_{\Rev}(w)\le k\}$.
\end{lem}
\begin{proof}
    The overall plan is to construct $\cB$ so that it captures the conditions in~\cref{lem: turning subwords shuffle}. Intuitively, $\cB$ keeps track of $k+1$ coordinates, each corresponding to a turning subword (that are nondeterministically constructed). The odd coordinates simulate the behavior of $\cA$, whereas the even ones simulate the reverse of $\cA$. 
    In addition $\cB$ checks (using its initial and accepting states) that the runs on the subwords can be correctly concatenated. We proceed with the precise details.

    Denote $\cA=\tup{\Sigma,Q,\delta,Q_0,F}$. We construct $\cB=\tup{\Sigma,Q',\delta',Q'_0,F'}$ as follows. $Q'=Q^{k+1}$, and the initial and final states are:
\begin{align*}
    Q'_0=\{(q_1, q_2, \ldots q_{k+1})\ |\ q_1\in Q_0\wedge q_i=q_{i+1} \text{ for all even } i\} \\
F'=\{(q_1, q_2, \ldots q_{k+1})\ |\ q_{k+1}\in F\wedge q_i=q_{i+1} \text{ for all odd } i\}
\end{align*}
For the transition function, we have that $(q'_0, q'_1,\ldots q'_k)\in\delta'((q_0, q_1,\ldots q_k), \sigma)$ if and only if there exists a single $1\le j\le k+1$ such that $q'_j\in \delta(q_j, \sigma)$ if $j$ is odd, and $q_j\in \delta(q'_j, \sigma)$ if $j$ is even. In addition, for every $i\neq j$, it holds that $q'_i=q_i$.

We turn to show the correctness of $\cB$. Consider an accepting run $\rho$ of $\cB$ on some word. Then $\rho$ starts at state $(q_1,q_2,\ldots,q_{k+1})\in Q'_0$ and ends at state $(s_1,s_2,\ldots,s_{k+1})\in F'$. By the definition of $\delta'$, we can split $\rho$ according to which component ``progresses'' in each transition, so that $\rho$ can be written as a shuffle of run $\rho^1,\ldots,\rho^{k+1}$ where $\rho^i$ leads from $q_i$ to $s_i$ in $\cA$ if $i$ is odd, and $\rho^i$ leads from $q_i$ to $s_i$ in the \emph{reverse} of $\cA$ if $i$ is even.
The latter is equivalent to $(\rho^i)^R$ (i.e., the reverse run of $\rho^i$) leading from $s_i$ to $q_i$ in $\cA$ if $i$ is even.

We now observe that these runs can be concatenated as follows: 
Recall that $q_1\in Q_0$ (by the definition of $Q'_0$). Then, $\rho^1$ leads from $q_1$ to $s_1$ in $\cA$. By the definition of $F$ we have $s_1=s_2$, and $(\rho^2)^R$ leads from $s_2$ to $q_2$ in $\cA$. Therefore, $\rho^1(\rho^2)^R$ leads from $q_1$ to $q_2$ in $\cA$. Continuing in the same fashion, we have $q_2=q_3$, and $\rho^3$ leading from $q_3$ to $s_3$, and so on up to $s_{k+1}$.

Thus, we have that $\rho^1(\rho^2)^R\rho_3\cdots (\rho^k)^R\rho^{k+1}$ is an accepting run of $\cA$. 

By identifying each accepting run $\rho^i$ with the subword it induces, we have that $w\in \regL(\cB)$ if and only if there are words $s_1,\ldots,s_{k+1}$ such that the two conditions in~\cref{lem: turning subwords shuffle} are satisfied.
\end{proof}
The proof of~\cref{lem:rev k construction} shows that the size of $\cB$ is polynomial in the size of $\cA$ and single-exponential in $k$, giving us $\cxPSPACE$ membership for \probkBoundedness, and $\cxEXPSPACE/$ $\cxEXPEXPSPACE$ for \probParamBoundedness with unary/binary encoding, respectively.

\subsection{\textsf{PSPACE-Hardness} of Boundedness for REV}
\label{sec:rev hardness}
Following a similar scheme to the Absolute Distance Semantics of~\cref{sec:absolute distance}, observe that for a word $w\in \Sigma^*$ we have that $\cA_\Rev(w)=0$ if and only if $w\in \regL(\cA)$, which implies that \probUnivBoundedness{$0$} is \cxPSPACEh. Yet again, the challenge is to prove hardness of \probkBoundedness for all values of $k$. 

\newcommand{\fsh}{\heartsuit}
\newcommand{\fss}{\spadesuit}

\begin{thm}
\label{thm:rev k boundedness PSPACE hard}
    For $\Rev$, \probkBoundedness is \cxPSPACEc for every $k\in \bbN$.
\end{thm}
\begin{proof}
Membership in $\cxPSPACE$ follows from~\cref{lem:rev k construction} and the discussion thereafter.
For hardness, we follow the same flow as the proof of~\cref{lem:dist k boundedness PSPACE hard}, but naturally the reduction itself is different. Specifically, we construct an \NFA that must read an expression of the form $(\fsh\fss)^m$ before its input. This allows us to shuffle the input to the form $\fss^m\fsh^m$, which causes many reversals. 

By~\cref{rmk:rev even} we can assume without loss of generality that $k$ is even, and we denote $k=2m$. 
We reduce the universality problem for $\NFAs$ to the \probUnivBoundedness{$2m$} problem. Consider an \NFA $\cA=\tup{Q,\Sigma,\delta,Q_0,\alpha}$, and let $\fsh,\fss\notin \Sigma$ be fresh symbols. 
We first check whether $\epsilon\in \regL(\cA)$. If $\epsilon\notin \regL(\cA)$, then $\regL(\cA)\neq \Sigma^*$ and we output some fixed unbounded automaton $\cB$ (e.g., as in~\cref{xmp:semantics ab simple,xmp:UUUU}).

Otherwise, we obtain from $\cA$ an \NFA $\cB$ over the alphabet $\Sigma\cup \{\fsh,\fss\}$ such that $w\in \regL(\cB)$ if and only if the following hold:
\begin{enumerate}
    \item Either $w$ does not contain exactly $m$ occurrences of $\fsh$ and of $\fss$, or
    \item $w=(\fsh\fss)^mx$ where $x\in \regL(\cA)$ (in particular $x\in \Sigma^*$).
\end{enumerate}
Constructing $\cB$ from $\cA$ is straightforward as the union of two components: one that accepts words that satisfy Condition 1 (using $2m+1$ states) and one for Condition $2$, which prepends to $\cA$ a component with $2m$ states accepting $(\fsh\fss)^m$. In particular, the reduction is in polynomial time.

We then have the following: if $\regL(\cA)=\Sigma^*$, then for every $w\in (\Sigma\cup \{\fsh,\fss\})^*$, if $w$ satisfies Condition 1, then $\cB_\Rev(w)=0$. Otherwise, $w$ has exactly $m$ occurrences of $\fsh$ and of $\fss$. Denote the indices of $\fsh$ by $i_1<i_2<\ldots<i_m$ and of $\fss$ by $j_1<j_2<\ldots<j_m$. Also denote by $t_0<t_1<\ldots<t_r$ the remaining indices of $w$. Then, consider the jump sequence 
\[
\vec{a}=(0,i_1,j_1,i_2,j_2,\ldots,i_m,j_m,t_0,t_1,\ldots,t_r,n+1)
\]
We claim that $w_\vec{a}\in \regL(\cB)$ by Condition 2. Indeed, $w$ starts with $(\fsh\fss)^m$, followed by letters in $\Sigma$ composing a word $x$. Since $x\in \regL(\cA)=\Sigma^*$, we have that Condition 2 holds. 
In addition, observe that since $t_0<t_1<\ldots<t_r<n+1$, then $\turnindices(\vec{a})\subseteq \{i_1,j_1,\ldots,i_m,j_m,t_0\}$, and in particular $|\turnindices(\vec{a})|\le 2m+1$. Moreover, by~\cref{rmk:rev even} we know that $|\turnindices(\vec{a})|$ is even, so in fact $|\turnindices(\vec{a})|\le 2m=k$. We conclude that $\cB_\Rev(w)\le k$, so $\cB$ is $k$-bounded.

Conversely, if $\regL(\cA)\neq \Sigma^*$, take $x\notin \regL(\cA)$ such that $x\neq \epsilon$ (which exists since we checked above that $\epsilon\in \regL(\cA)$). Consider the word $w=\fss^m\fsh^m x$, then have $w\notin \regL(\cB)$. We claim that $\cB_\Rev(w)>2m$. Indeed, if there exists $\vec{a}$ such that $w_\vec{a}\in \regL(\cB)$, then since $w$ has exactly $m$ occurrences of $\fss$ and of $\fsh$, it must be accepted by Condition 2. By the structure of $w$, the jump sequence $\vec{a}$ needs to permute $\fss^m\fsh^m$ into $(\fsh\fss)^m$. Intuitively, this means that the head must jump ``back and forth'' for $2m$ steps. More precisely, for every $i\in \{1,\ldots, |w|\}$ it holds that
\[a_i\in \begin{cases}
    \{m+1,\ldots,2m\} & i\le 2m\text{ is odd}\\
    \{1,\ldots,m\} & i\le 2m\text{ is even}\\
    \{2m+1,\ldots,|w|\} & i>2m\\
\end{cases}\]
In particular, $\{1,\ldots,2m\}\subseteq \turnindices(\vec{a})$.
Observe that the remaining suffix of $w_\vec{a}$ starting at $2m+1$ cannot be $x$, since $x\notin \regL(\cA)$, so $\vec{a}$ is not the identity starting from $2m+1$. It therefore has an additional reversal cost of at least $1$. Thus, $|\turnindices(\vec{a})|>2m$. In particular, $\cB_{\Rev}(w)>2m$, so $\cB$ is not $2m$-bounded, and we are done.
\end{proof}
As in~\cref{sec:abs hardness}, it follows that \probParamBoundedness,\ \probDomkBoundedness and \probDomParamBoundedness are also $\cxPSPACEh$.

\section{The Hamming Semantics}
\label{sec:hamming}
Recall from~\cref{def:hamming semantics} that for a \JFA $\cA$ and word $w$, the cost $\cA_{\Ham}(w)$ is the minimal Hamming distance between $w$ and $w'$ where $w'\sim w$ and $w'\in \regL(\cA)$.
\begin{rem}[An alternative interpretation of the Hamming Semantics]
    We can think of a jumping automaton as accepting a permutation $w'$ of the input word $w$. 
    As such, a natural candidate for a quantitative measure is the ``distance'' of the permutation used to obtain $w'$ from the identity (i.e. from $w$). The standard definition for such a distance is the number of transpositions of two indices required to move from one permutation to the other (this is commonly known as the  distance in the Cayley graph~\cite{magnus2004combinatorial} for the transpositions generators of $S_n$). It is easy to show that in fact, the Hamming distance coincides with this definition.
\end{rem}

Again we start by establishing the complexity of computing the Hamming measure of a given word.
\begin{thm}[Hamming \probMembership is $\cxNPc$]
\label{thm:ham membership NPc}
The problem of deciding, given $\cA$ and $k\in \bbN$, whether $\cA_{\Ham}(w)\le k$ is $\cxNPc$.
\end{thm}
\begin{proof}
By definition we have that $\cA_{\Ham}(w)\le |w|$ for every word $w$. Thus, in order to decide whether $\cA_{\Ham}(w)\le k$ we can nondeterministically guess a permutation $w'\sim w$ and verify that $w'\in \regL(\cA)$ and that $d_H(w,w')\le k$. Both conditions are computable in polynomial time. Therefore, the problem is in $\cxNP$.

Hardness follows (similarly to the proof of~\cref{thm:dist membership NPcomplete}) by reduction from membership in \JFA, noting that $w\in \jL(\cA)$ if and only if $\cA_{\Ham}(w)\le |w|$.
\end{proof}

Similarly to~\cref{sec:abs decidability,sec:rev boundedness decidable}, in order to establish the decidability of \probParamBoundedness, we start by constructing an \NFA that accepts exactly the words for which $\cA_{\Ham}(w)\le k$.
\begin{lem}
\label{lem:ham NFA construction}
    Let $k\in \bbN$. We can effectively construct an \NFA $\cB$ with $\regL(\cB)=\{w\in \Sigma^*\mid \cA_\Ham(w)\le k\}$.
\end{lem}
\begin{proof}
    Let $k\in \bbN$. Intuitively, $\cB$ works as follows: while reading a word $w$ sequentially, it simulates the run of $\cA$, but allows $\cA$ to intuitively ``swap'' the current letter with a (nondeterministically chosen) different one (e.g., the current letter may be $a$ but the run of $\cA$ can be simulated on either $a$ or $b$). Then, $\cB$ keeps track of the swaps made by counting for each letter $a$ how many times it was swapped by another letter, and how many times another letter was swapped to it. This is done by keeping a counter ranging from $-k$ to $k$, counting the difference between the number of occurrences of each letter in the simulated word versus the actual word. We refer to this value as the \emph{balance} of the letter.
    $\cB$ also keeps track of the total number of swaps.
    Then, a run is accepting if at the end of the simulation, the total amount of swaps does not exceed $k$, and if all the letters end up with $0$ balance.
    
    We now turn to the formal details. Recall that $\cA=\tup{\Sigma,Q,\delta,Q_0,\alpha}$. We define $\cB=\tup{\Sigma,Q',\delta',Q_0',\beta}$.

    The state space of $\cB$ is $Q'=Q\times \{-k,\ldots,k\}^\Sigma\times \{0,\ldots,k\}$. We denote a state of $\cB$ by $(q,f,c)$ where $q\in Q$ is the current state of $\cA$, $f:\Sigma\to \{-k,\ldots,k\}$ describes for each letter its \emph{balance} and $c\in \{0,\ldots,k\}$ is the total number of swaps thus far.

    The initial states of $\cB$ are $Q'_0=\{(q,f,0)\mid q\in Q_0\wedge f(\sigma)=0 \text{ for all } \sigma\in \Sigma\}$. That is, we start in an initial state of $\cA$ with balance and total cost of $0$.
    The transition function is defined as follows. Consider a state $(q,f,c)$ and a letter $\sigma\in \Sigma$, then $(q',f',c')\in \delta'((q,f,c),\sigma)$ if and only if either $q'=\delta(q,\sigma)$ and $f'=f$ and $c'=c$, or there exists $\tau\in \Sigma$, $\tau\neq \sigma$ such that $q'\in \delta(q,\tau)$, $c'=c+1$, $f'(\sigma)=f(\sigma)-1$, and $f'(\tau)=f(\tau)+1$. That is, in each transition we either read the current letter $\sigma$, or swap for a letter $\tau$ and update the balances accordingly.

    Finally, the accepting states of $\cB$ are $\beta=\{(q,f,c)\mid q\in \alpha\wedge f(\sigma)=0 \text{ for all }\sigma\in \Sigma\}$.

    In order to establish correctness, we observe that every run of $\cB$ on a word $w$ induces a word $w'$ (with the nondeterministically guessed letters) such that along the run the components $f$ and $c$ of the states track the swaps made between $w$ and $w'$. In particular, $c$ keeps track of the number of total swaps, and $\sum_{\sigma\in \Sigma}f(\sigma)=0$. Moreover, for every word $\sigma$, the value $f(\sigma)$ is exactly the number of times $\sigma$ was read in $w'$ minus the number of times $\sigma$ was read in $w$. 

    Since $\cB$ accepts a word only if $f\equiv 0$ at the last state, it follows that $\cB$ accepts if and only if $w'\sim w$, and the run of $\cA$ on $w'$ is accepting. Finally, since $c$ is bounded by $k$ and is increased upon every swap, then $w$ is accepted if and only if its cost is at most $k$. Note that since $c$ is increased upon each swap, then limiting the image of $f$ to values in $\{-k,\ldots, k\}$ does not pose a restriction, as they cannot go beyond these bounds without $c$ going beyond the bound $k$ as well.
\end{proof}
An analogous proof to~\cref{thm:abs decidable boundedness} gives us the following.
\begin{thm}
\label{thm:ham decidable boundedness}
The following problems are decidable for the $\Ham$ semantics: \probDomkBoundedness,\  \probDomParamBoundedness,\ \probkBoundedness, and \probParamBoundedness.
\end{thm}
We note that the size of $\cB$ constructed in~\cref{lem:ham NFA construction} is polynomial in $k$ and single-exponential in $|\Sigma|$, and therefore when $\Sigma$ is fixed and $k$ is either fixed or given in unary, both \probParamBoundedness and \probkBoundedness are in $\cxPSPACE$. 

For a lower bound, we remark that similarly to~\cref{sec:abs hardness}, it is not hard to prove that \probkBoundedness is also $\cxPSPACEh$ for every $k$, using relatively similar tricks. However, since \probParamBoundedness is already $\cxPSPACEc$, then \probkBoundedness is somewhat redundant. We therefore make do with the trivial lower bound whereby we reduce universality of $\NFA$ to \probUnivBoundedness{$0$}. 
\begin{thm}
\label{thm:ham fixed boundedness decidable}
For $\Ham$, the \probParamBoundedness problem is $\cxPSPACEc$ for $k$ encoded in unary and fixed alphabet $\Sigma$.
\end{thm}

\section{The Maximum Jump Semantics}
\label{sec:maxjump}
Of all the semantics we propose, the $\Max$ semantics turns out to be the most technically-challenging to analyze. 
We start with a fundamental observation regarding the structure of jump sequences that have a bounded $\Max$ semantics. 
Consider a jump sequence $\vec{a}$ such that $\sem{\vec{a}}_{\Max}\le k$ for some $k$. It is not hard to see that if $\vec{a}$ is very long, the head may perform arbitrarily many reversals during $\vec{a}$ (i.e., the number of reversals can be arbitrarily larger than $k$), e.g., by repeating sequences of indices of the form $(i,i+2,i+1,i+3)$.
However, if we fix a concrete index $j$ on the tape, then $\vec{a}$ cannot cross $j$ too many times. Indeed, after enough crosses, all cells within distance $k$ of $j$ have been read, and therefore in order to cross $j$ again a jump of size $>k$ needs to be made.

This observation provides some structures to $\Max$-bounded sequences. We now start by formalizing it. We reuse some notions form the Reversal semantics in \cref{sec:reversal}. 
Our first definition is a variant of \emph{turning subwords}, but we consider indices instead of words, and we divide them to left and right.
\begin{defi}[Sweeps]
\label{def:sweep}
    Let $\vec{a}=(a_0,\ldots,a_{n+1})$ be a jump sequence. A \emph{right sweep} of $\vec{a}$ is an infix $\vec{b}=(a_i,\ldots,a_{i+j})$ such that all of the following hold:
    \begin{itemize}
        \item $a_k<a_{k+1}$ for all $i\leq k<i+j$.
        \item Either $i=0$, or $i>1$ and $a_{i-2}>a_{i-1}$.
        \item Either $i+j=n+1$ or $a_{i+j+1}<a_{i+j}$.
    \end{itemize}
    Similarly, a \emph{left sweep}, is an infix $\vec{b}=(a_i,\ldots,a_{i+j})$ such that:
    \begin{itemize}
        \item $a_k>a_{k+1}$ for all $i\le k<i+j$.
        \item $i>1$ and $a_{i-2}<a_{i-1}$.
        \item $i+j<n+1$ and $a_{i+j+1}>a_{i+j}$.
    \end{itemize}
    We use \emph{sweep} to refer to either a right or a left sweep.
    
    Equivalently, $\vec{b}$ is a sweep if:
    \begin{itemize}
        \item Either $i=0$ or $i-1\in\turnindices(\vec{a})$.
        \item $i+j\in\turnindices({\vec{a}})$.
        \item $k\notin\turnindices({\vec{a}})$ for all $i\le k<i+j$.
    \end{itemize}
\end{defi}
\noindent 
Note that every jump sequence can be written uniquely as a concatenation of sweeps.

Next, we consider the partition of the tape to two parts at some index, and the number of times the border of the partition is crossed.
\begin{defi}
    \label{def:m-cut and k-crossing}
    Let $n\in\bbN$ and $0\le m\le n$. The \emph{$m$-cut} of $\{0,\ldots,n+1\}$ is the partition $(\{0,\ldots,m\},\{m+1,\ldots,n+1\})$.
    
    Let $\vec{a}=(a_0,\ldots,a_{n+1})$ be a jump sequence. We say that \emph{$\vec{a}$ $k$-crosses the $m$-cut} if there exist distinct indices $i_1,\ldots,i_k\in \{0,\ldots,n\}$ such that for all $1\le j\le k$, $a_{i_j},a_{i_j+1}$ belong to two different sides of the $m$-cut.

    We denote the maximal $k$ such that $\vec{a}$ $k$-crosses the $m$-cut by $\numcross(\vec{a},m)$.
\end{defi}
\noindent 
Our fundamental observation is that if $\sem{\vec{a}}_{\Max}$ is bounded, then so is the maximal crossing.
\begin{lem}
    \label{lem:bounded max => bounded cut-crossing}
    Let $\vec{a}$ be a jump sequence of length $n$. If $\sem{\vec{a}}_\Max\le k$, then $\numcross(\vec{a},m)\le 2k+1$ for all $1\le m\le n$.
\end{lem}
\begin{proof}
    Intuitively, since the maximal jump is bounded by $k$, and since no index can be visited twice, then after $2k+1$ iterations all the cells within distance $k$ of $m$ have been visited, and there is no way to jump over $m$ again while maintaining a maximal jump of at most $k$. The following argument formalizes this intuition.

    Let $0\le i_1,\ldots,i_{\numcross(\vec{a},m)}\le n$ be the indices provided by \cref{def:m-cut and k-crossing}. For all $1\le j\le \numcross(\vec{a},m)$ we have that either $a_{i_j}\le m<a_{i_j+1}$ or $a_{i_j}\le m<a_{i_j+1}$. Additionally, $\sem{a_{i_j+1}-a_{i_j}}\le k$. It follows that $a_{i_j}\in\{m-k,m-k+1,\ldots,m+k+1\}$. Since $a_{i_j}$ are all distinct, we have $\numcross(\vec{a},m)\le|\{m-k,\ldots,m+k+1\}|=2k+2$. Observe that $\numcross(\vec{a},m)$ is odd (since $\vec{a}$ begins at index $0$ and ends at $n+1$), and so $\numcross(\vec{a},m)\le2k+1$. 
\end{proof}

\begin{lem}
\label{lem:max NFA construction}
    Let $k\in \bbN$. We can effectively construct an \NFA $\cB$ with $\regL(\cB)=\{w\in \Sigma^*\mid \cA_\Max(w)\le k\}$.
\end{lem}
We prove the lemma in the remainder of this section. 
We first provide some intuition behind the construction of $\cB$. 
Similarly to the construction in \cref{lem:rev k construction}, $\cB$ keeps track of a fixed number (depending on $k$) of coordinates, each can be ``active'' or ``inactive'' at any given time, and when active, corresponds to a sweep. 
Unlike the $\Rev$ semantics, however, it is now not the case that the number of sweeps is bounded, so we cannot a-priori assign a coordinate to each sweep.

In order to overcome this obstacle, we observe that by \cref{lem:bounded max => bounded cut-crossing}, we have a bound of the number of active sweeps that need to be tracked at any given time. $\cB$ then assigns every letter it reads to one of the active sweeps, and simulates the behavior of $\cA$ or the reverse of $\cA$ on each right or left sweep respectively. At each step, $\cB$ might ``activate'' a pair of inactive coordinates, thus assigning one to a left sweep ending in some state $q$, and a subsequent right sweep starting in $q$. Similarly, $\cB$ might ``deactivate'' a pair of active coordinates to mark the end of a right sweep (at a state $q$) and the beginning of the subsequent left sweep (at the same state $q$). The set of different ``operations'' that can be performed by $\cB$ is captured by the following definitions.

This intuition gets us close to the solution, but not quite there due to the fact that the operations sequences we allow cannot always be translated to runs. Below we formally define the operations, and demonstrate this issue. We then proceed to resolve it and give the complete construction.

\begin{defi}
\label{def:operations (max semantics)}
    Let $k\in\bbN$. The \emph{$k$-operations} are the members of the set
    \begin{align*}
        & \{(\activate,j_L,j_R)\mid 1\le j_L,j_R\le\numslots\}
        \cup \\
        & \{(\deactivate,j_L,j_R)\mid 1\le j_L,j_R\le\numslots\}
        \cup \\
        & \{\none(j)\mid 1\le j\le\numslots\}
    \end{align*}
    We sometimes omit $k$ when it is clear from context.
\end{defi}
Intuitively, $\none(j)$ corresponds to not activating or deactivating, and reading a letter into coordinate $j$. While performing $\activate(j_L,j_R)$ we must read a letter into $j_L$, and while performing $\deactivate(j_L,j_R)$ we must read a letter into $j_R$.

\begin{defi}
\label{def:op sequences}
    Let $\vec{u}=(\vec{u}_1,\ldots,\vec{u}_n)$ be a sequence of $k$-operations. For $1\le j\le\numslots,1\le i \le n$ we say that \emph{$j$ is right-active at step $i$} if there exists $i'\le i$ such that $\vec{u}_{i'}$ is of the form $(\activate,j',j)$, and for all $i'<i''<i$, $\vec{u}_{i''}$ is not a deactivation involving $j$. Note that $i$ itself is allowed to be the deactivation of $j$.
    We similarly define $j$ being \emph{left-active} at step $i$ if there exists $i'\le i$ such that $\vec{u}_{i'}$ is of the form $(\activate,j,j')$, and for all $i'<i''<i$, $\vec{u}_{i''}$ is not a deactivation involving $j$.
    
    If $j$ is right-active or left-active at step $i$, then it is \emph{active} at step $i$, and otherwise \emph{inactive}. Additionally we say that $1$ is right-active at step $0$.
    
    Naturally, not all sequences of $k$-operations are ``legal''. Thus, $\vec{u}$ is a \emph{$k$-operation sequence} if for all $1\le i\le n$:
    \begin{itemize}
        \item If $\vec{u}_i=(\activate,j_L,j_R)$ then $j_L,j_R$ are both inactive at step $i-1$.
        \item If $\vec{u}_i=(\deactivate,j_L,j_R)$ then $j_L,j_R$ are respectively left-active and right-active at step $i-1$.
        \item If $\vec{u}_i=(\none,j)$ then $j$ is active at step $i$.
        \item At step $n$, there exists a single $j$ that is right-active, and all other coordinates are inactive.
    \end{itemize}
    If $j$ is being deactivated at step $i$, we call it \emph{weakly active} at step $i$. Otherwise, if it is active at step $i$, it is \emph{strongly active} at step $i$.
\end{defi}

In the following, we keep track of each sweep in some coordinate. In order to allow a fixed number of coordinates, we present the following definitions and lemmas.

\begin{defi}
\label{def:sweep range}
    Let $\vec{a}$ be a jump sequence and let $\vec{b}=(a_i,\ldots,a_{i+j})$ be a right sweep. The \emph{range} of $\vec{b}$ is:
    \begin{itemize}
        \item If $i=0$, then $\range(\vec{b})=\{a_i,a_i+1\ldots,a_{i+j}\}$.
        \item If $i\neq0$, then $\range(\vec{b})=\{a_{i-1},a_{i-1}+1,\ldots,a_{i+j}\}$.
    \end{itemize}
    Similarly, if $\vec{b}$ is a left sweep, we define $\range(\vec{b})=\{a_{i+j},a_{i+j}+1,\ldots,a_{i-1}\}$. 
\end{defi}
The distinction in the right-sweep case is due to the fact that in the case that $\vec{b}$ comes after a left sweep $\vec{b'}$, we want the range to include all indices starting at the turning index, while $\vec{b}$ only starts further to the right. For left sweeps, cases are not needed since a left sweep is never the first sweep.

The bounded crossing property can be cast in terms of ranges of sweeps, as follows.
\begin{lem}
\label{lem:max number of simultaneous sweeps}
    Consider a jump sequence $\vec{a}=(a_0,\ldots,a_{n+1})$ with $\sem{\vec{a}}_\Max=k$. 
    For every $i\in \{1,\ldots, n\}$ define 
    $R_i=\{\vec{b}\mid\vec{b}\text{ is a sweep of $\vec{a}$ and }i\in\range(\vec{b})\}$, then $|R_i|\le \numslots$.
\end{lem}
\begin{proof}
    Let $0\le i\le n$. There exists at most one sweep in $R_i$ containing $i$ (as there exists exactly one such sweep overall). Let $\vec{b}\in R_i$ be a sweep not containing $i$. We claim that there exists an index $j_\vec{b}$ such that $a_{j_\vec{b}}\in\vec{b}$ and $a_{j_\vec{b}},a_{j_\vec{b}+1}$ belong to two different sides of the $i$-cut. Indeed, write $\vec{b}=(a_{i'},\ldots,a_{i'+j'})$, and observe that $j'>0$ (as otherwise $\vec{b}=(i)$ and in particular it contains $i$). From $i\in\range(\vec{b})$ it follows that $a_{i'},a_{i'+j'}$ belong to different sides of the $i$-cut. Let $j_\vec{b}<i'+j'$ be the maximal such that $a_{j_\vec{b}}$ belongs to the same side as $a_{i'}$. Then $j_\vec{b}$ satisfies the condition.

    Since the sweeps are disjoint, the indices $j_\vec{b}$ are distinct, and so by \cref{def:m-cut and k-crossing}, $\vec{a}$ $(|R_i|-1)$-crosses the $i$-cut. It follows by \cref{lem:bounded max => bounded cut-crossing} that $|R_i|\le(2k+1)+1=\numslots$.
\end{proof}

We now consider the following setting: we allocate $2k+1$ coordinates, and upon reading a word (sequentially), we wish to track each ``active'' sweep in a coordinate. We present some definitions to establish these notions.
\begin{defi}
\label{def:embedding}
    Let $k,n\in\bbN$ and let $P$ be a partition of $\{1,\ldots, n\}$. A \emph{$k$-embedding} of $P$ is a function $f\colon P\to\{1,\ldots,\numslots\}$.
\end{defi}

\begin{defi}
\label{def:partition induced by jump sequence}
\label{def:valid embedding for jump sequence}
    Let $\vec{a}=(\vec{a}_0,\ldots,\vec{a}_{n+1})$ be a jump sequence and let $\vec{b}_1,\ldots,\vec{b}_m$ be the sweeps of $\vec{a}$. The \emph{partition of $\{1,\ldots, n\}$ induced by $\vec{a}$} is $P_\vec{a}=\{B_1,\ldots B_m\}$ such that $B_i$ is the set of indices appearing in $\vec{b}_i$ for all $i\in\{1,\ldots, m\}$.

    For $k\in \bbN$, a $k$-embedding $f$ of $P_\vec{a}$ is called \emph{valid for $\vec{a}$} if for every two sweeps $\vec{b},\vec{b}'$ with respective $B,B'\in P_\vec{a}$, if $f(B)=f(B')$ then $\range(\vec{b})\cap\range(\vec{b}')=\emptyset$.
\end{defi}
Intuitively, this definition says that if two sweeps are assigned to the same coordinate, then they need to have disjoint ranges, i.e., one of them completes before the other starts.


\begin{lem}
    \label{lem:existence of valid embedding for jump sequence}
    Let $\vec{a}$ such that $\sem{\vec{a}}_\Max\le k$. Then $\vec{a}$ has a valid $k$-embedding.
\end{lem}
\begin{proof}
    For each sweep $\vec{b}$ of $\vec{a}$, let $i_\vec{b}$ be the minimal index in $\range(\vec{b})$ (ties broken arbitrarily). Let $\vec{b}_1\le\ldots\le\vec{b}_m$ be the sweeps, ordered by $i_\vec{b}$. We assign the values of $f$ to the sweeps in increasing order, using a greedy algorithm: When assigning $f(B)$ for $\vec{b}$, applying \cref{lem:max number of simultaneous sweeps} to $i_\vec{b}$, there is at least one free value, and we choose one arbitrarily.
\end{proof}

A valid $k$-embedding essentially ``dictates'' which sweeps should be assigned to which coordinates. 
Specifically, we have the following.
\begin{defi}
\label{def:op sequence of jump sequence and embedding}
    Let $k\in\bbN$, Let $\vec{a}=(\vec{a}_0,\ldots,\vec{a}_{n+1})$ be a jump sequence and let $f$ be a valid $k$-embedding of $\vec{a}$. Let $\vec{b}_1,\ldots,\vec{b}_m$ be the sweeps of $\vec{a}$ and $P_\vec{a}=\{B_1,\ldots,B_m\}$ the induced partition of $\{1,\ldots,n\}$. The \emph{operation sequence of $(\vec{a},f)$} is the unique operation sequence $\vec{u}^{\vec{a},f}$ defined as follows. For all $i\in\{1,\ldots,n\}$:
    \begin{itemize}
        \item If $i$ is the minimal index of a left sweep $\vec{b}_l$, then either $i=0$, or \[\vec{u}^{\vec{a},f}_i=(\activate,f(B_l),f(B_{l+1}))\]
        \item If $i$ is the maximal index of a right sweep $\vec{b}_l$, then either $i=n+1$, or \[\vec{u}^{\vec{a},f}_i=(\deactivate,f(B_l),f(B_{l+1}))\]
        \item Otherwise, $\vec{u}^{\vec{a},f}_i=(\none,f(B))$ such that $i\in B$.
    \end{itemize}
\end{defi}
\noindent 
Dually, any operation sequence induces a partition and a $k$-embedding, as follows.
\begin{defi}
\label{def:partition and embedding of op sequence}
    Let $k\in\bbN$ and $\vec{u}$ an operation sequence of length $n$. The \emph{partition and embedding of $\vec{u}$} are the partition $P_\vec{u}$ of $\{1,\ldots,n\}$ and $k$-embedding $f_\vec{u}$ are defined as follows. Let $f'\colon\{1,\ldots,n\}\to\{1,\ldots,\numslots\}$ be
    \[
    f'(i)=\begin{cases}
        j_L & \vec{u}_i=(\activate,j_L,j_R) \\
        j_R & \vec{u}_i=(\deactivate,j_L,j_R) \\
        j & \vec{u}_i=(\none,j) \\
    \end{cases}
    \]
    $P_\vec{u}$ is defined by the equivalence relation where for $i<i'$, $i\sim i'$ if $f'(i)=f'(i')$ and there does not exist $i<i''<i'$ such that $\vec{u}_{i''}$ is activation or deactivation involving $f'(i)$. $f_\vec{u}$ is defined to be $f_\vec{u}(B)=f'(i)$ for any $i\in B$.
\end{defi}
It is tempting to think that $k$-embeddings and operation sequences are just two representations of the same objects. However, things are somewhat more intricate.
Note that if $\vec{a}$ is a jump sequence and $f$ is a valid embedding for it
then $P_{\vec{u}^{\vec{a},f}},f_{\vec{u}^{\vec{a},f}}$ are exactly $P_\vec{a},f$. 
However, not every operation sequence $\vec{u}$ is obtained as $\vec{u}^{\vec{a},f}$ for some $\vec{a},f$. Intuitively, this happens when the operation sequence does not ``connect'' the sweeps to a single run, as we now demonstrate.
\begin{exa}
\label{xmp:op sequence with a cycle}    
    Consider $k=2$, $n=4$, and the sequence
    \[\vec{u}=((\none,1),(\activate,2,3),(\deactivate,2,3),(\none,1)\] Coordinate $1$ is active and holds a single ``sweep'' through the run, into which we read $w_1,w_4$. Additionally, coordinates $2,3$ are activated at step $2$ and we read $w_2$ into $2$, then we read $w_3$ into $3$ and deactivate $2,3$. The rest of the coordinates are never active. However, there is no fitting jump sequence, and intuitively the reason is the ``sweeps'' do not form a single path with their activations and deactivations.

    Moreover, the issue raised in this example is not something that can be fixed ``locally'' in the definition. Indeed, upon reading the prefix $(\none,1),(\activate,2,3)$ of $\vec{u}$, there is no way of telling whether the suffix that is to be read induces a legal jump sequence or not.
\end{exa}

We lift \cref{def:sweep range} to operation sequences that do not necessarily stem from jump sequences. Intuitively, this involves defining the range of the partitions, which we do by taking the minimal interval between activation and de-activation, as follows.
\begin{defi}
    \label{def:op sequence partition range}
    Let $\vec{u}$ be a $k$-operation sequence and $P,f$ the partition and embedding of $\vec{u}$. For $B\in P$, the \emph{range of $B$ induced by $\vec{u}$}, is $\range_\vec{u}(B)=\{i_1,\ldots,i_2\}$, where $i_1,i_2$ are defined as follows. Choose $i\in B$ arbitrarily.
    \begin{itemize}
        \item If $f(B)$ is activated at some step $i'\le i$, then $i_1$ is the maximal such $i'$, and otherwise $i_1=0$.
        \item If $f(B)$ is deactivated at some step $i'\ge i$, then $i_2$ is the minimal such $i'$, and otherwise $i_2=n+1$.
    \end{itemize}
\end{defi}
\noindent 
Note that the ranges above are well-defined (i.e., independent of the choice of $i$) due to the equivalence relation defined in \cref{def:partition and embedding of op sequence}.
We additionally define an ``inverse'' of the embedding, which tells us to which part in the partition each coordinate/index pair belongs to, as follows.
\begin{defi}
\label{def:embedding inverse}
    Let $\vec{u}$ be a $k$-operation sequence and $P,f$ the partition and embedding of $\vec{u}$. The \emph{inverse embedding} is the function $f^{-1}\colon\{1,\ldots,\numslots\}\times\{0,\ldots,n+1\}\to P$ (by abuse of notation) such that $f^{-1}(j,i)$ is the unique $B\in P$ such that $f(B)=j$ and $i\in\range(B)$, or $\bot$ if no such $B$ exists.
\end{defi}

As mentioned above, the problematic operations sequences are those whose embeddings do not correspond to a single run. As demonstrated in \cref{xmp:op sequence with a cycle}, this happens when there are disconnected components. We now formalize this property by considering an appropriate graph.
\begin{defi}
\label{def:sweep graph}
    Let $\vec{u}$ be a $k$-operation sequence and $P,f$ the partition and embedding of $\vec{u}$. The \emph{sweep graph of $\vec{u}$} is $G_\vec{u}=(V,E)$ defined as follows. The vertices are $V=P$, and two sets in $P$ are connected by an edge if they are activated or deactivated as a pair. That is, $(B,B')\in E$ if there exists $i\in\range(B)\cap\range(B')$ such that $\vec{u}_i$ is an activation or deactivation of $f(B),f(B')$.
\end{defi}
Observe that the degree of the vertices in $G$ is at most $2$, since every sweep is activated and deactivated at most once. Thus, $G$ is a disjoint union of simple paths and simple cycles.
Intuitively, $G$ is of interest to us because its connectivity is equivalent to $\vec{u}$ stemming from a jump sequence, and avoiding cases like \cref{xmp:op sequence with a cycle}. Therefore, the NFA we construct keeps track of the connected components of $G$ during the run. The equivalence is captured in the following lemma.

\begin{lem}
\label{lem:G is a path <-> there exists a jump sequence}
    Let $\vec{u}$ be a $k$-operation sequence and $P,f$ the partition and embedding of $\vec{u}$. There exists a jump sequence $\vec{a}$ such that $\vec{u}=\vec{u}^{\vec{a},f}$ if and only if $G_\vec{u}$ is a simple path.
\end{lem}
\begin{proof}
    Assume $G_\vec{u}$ is a path $B_{i_1},\ldots,B_{i_m}$. For all $1\le l\le m$ let $\vec{b}_{i_l}$ consist of the elements of $B_{i_l}$ in increasing order if $l$ is odd, and in decreasing order if $l$ is even. Then $\vec{a}=(0,\vec{b}_{i_1},\ldots,\vec{b}_{i_m},n+1)$ satisfies the requirement.
    
    For the other direction, assume there exists such $\vec{a}$. Then the edges of $G_\vec{u}$ correspond to exactly the pairs of subsequent sweeps of $\vec{a}$, with every edge connecting a sweep to the next one, and so it is a path.
\end{proof}
In order to reason about the components of the graph above in a ``sequential'' manner, we define a function that indicates for every coordinate/index pair, which are the sweeps that are connected to this coordinate. The precise definition is the following.
\begin{defi}
\label{def:component function}
    Let $\vec{u}$ be a $k$-operation sequence. The \emph{component function of $\vec{u}$} is the function $S\colon(\{1,\ldots,\numslots\}\times\{0,\ldots,n\})\to 2^{\{1,\ldots,\numslots\}}\cup\{\bot\}$ defined inductively as follows. 
    \begin{itemize}
        \item $S(1,0)=\{1\}$ and $S(j,0)=\bot$ for all $j\neq1$.
        \item If $i>0$ and $\vec{u}_i=(\activate,j_L,j_R)$ then:
        \begin{itemize}
            \item $S(j_L,i)=S(j_R,i)=\{j_L,j_R\}$.
            \item $S(j,i)=S(j,i-1)$ for all $j\notin\{j_L,j_R\}$.
        \end{itemize}
        \item If $i>0$ and $\vec{u}_i=(\deactivate,j_L,j_R)$ then:
        \begin{itemize}
            \item $S(j_L,i)=S(j_R,i)=\bot$.
            \item $S(j,i)=S(j_L,i-1)\cup S(j_R,i-1)$ for all $j\in(S(j_L,i-1)\cup S(j_R,i-1))\setminus\{j_L,j_R\}$.
            \item $S(j,i)=S(j,i-1)$ for all $j\notin S(j_L,i-1)\cup S(j_R,i-1)$.
        \end{itemize}
        \item If $i>0$ and $\vec{u}_i=(\none,j)$ then $S(j',i)=S(j',i-1)$ for all $j'$.
    \end{itemize}
\end{defi}
\noindent 
Note that for $j\neq1$, if $j$ is active at step $i$ then $f^{-1}(j,i)\neq\bot$. For $j=1$, however, since it starts as active, if it is never ``used'' by $u$ then it remains active through $\vec{u}$ but there is no $B$ with $f(B)=1$. Hence, we add the restriction that $1$ must be used.
\begin{defi}
    \label{def:1-valid k-operation}
    Let $\vec{u}$ be a $k$-operation sequence. We say that $\vec{u}$ is \emph{$1$-valid} if $\vec{u}$ contains an operation involving the coordinate $1$.
\end{defi}
Observe that if $\vec{u}$ is $1$-valid then $f^{-1}(1,0)$ is the set $B$ containing the minimal index $i$ such that $\vec{u}_i$ involves $1$.

We can now characterize the existence of a cycle (i.e., an isolated connected component) in $G_\vec{u}$ in a sequential manner. Intuitively, the characterization is that at some point we deactivate two coordinates that are connected \emph{only} to each other. Since the structure of a cycle may involve several sweeps, the proof is somewhat involved.

\begin{lem}
\label{lem:sweep graph cycle characterization}
    Let $\vec{u}$ be a $1$-valid $k$-operation sequence. Then $G_\vec{u}$    
    has a cycle if and only if there exist $i>0,j_L,j_R$ such that $\vec{u}_i=(\deactivate,j_L,j_R)$ and $S(j_L,i-1)=S(j_R,i-1)=\{j_L,j_R\}$.
\end{lem}
\begin{proof}
    Let $P,f$ be the partition and embedding of $\vec{u}$. For all $i\in\{0,\ldots,n\}$, Let $G_i=(V_i,E_i)$ be the sub-graph of $G_\vec{u}$ defined as follows. $V_i=\{B\in P\mid \exists i'\le i,i'\in\range_\vec{u}(B)\}$.
    Intuitively, $V_i$ is the set of all sets that are activated up until (and including) step $i$. $E_i$ contains all the pairs $(B,B')$ such that there exists $i'\le i,i'\in\range(B)\cap\range(B')$ such that $\vec{u}_{i'}$ is activation or deactivation involving $f(B),f(B')$. Intuitively, $B,B'$ are activated or deactivated together up until (including) step $i$. 
    We claim that for all $j\in\{1,\ldots,\numslots\},i\in\{0,\ldots,n\}$:
    
    \begin{itemize}
        \item If $j$ is inactive at step $i$, then $S(j,i)=\bot$.
        \item If $j$ is weakly active at step $i$, then $S(j,i)=\bot$.
        \item Otherwise ($j$ is strongly active at step $i$), $S(j,i)$ is the set of coordinates in $f(C)$ that are strongly active at step $i$, where $C$ is the connected component of $f^{-1}(j,i)$ in $G_i$.
    \end{itemize}
    The proof is by induction on $i$. For $i=0$, only $j=1$ is (strongly) active, $G_0$ consists of the single vertex $f^{-1}(1,0)$, and indeed we have $S(1,0)=\{1\}$ and $S(j,0)=\bot$ for all $j\neq1$ by definition. Let $i>0$ and assume the claim holds for $i-1$. We consider three cases according to the $i$'th operation:
    \begin{itemize}
        \item $\vec{u}_i$ is of the form $(\none,j)$. Then the claim follows from the fact that $G_{i+1}=G_i$, $S(j',i)=S(j',i-1)$ for all $j'$, and the induction hypothesis.
        \item $\vec{u}_i$ is of the form $(\activate,j_L,j_R)$.     
        Then 
        \[V_i=V_{i-1}\cup\{f^{-1}(j_L,i),f^{-1}(j_R,i)\} \text{ and } E_i=E_{i-1}\cup\{(f^{-1}(j_L,i),f^{-1}(j_R,i))\}\] 
        Hence, The connected components of $G_i$ are the same as those of $G_{i-1}$ with the addition of $\{f^{-1}(j_L,i),f^{-1}(j_R,i)\}$. By the definition of $S$ we have $S(j_L,i)=S(j_R,i)=\{j_L,j_R\}$ and $S(j,i)=S(j,i-1)$ for all other $j$. If $j$ is inactive in step $i$ then $j\notin\{j_L,j_R\}$ and it is also inactive in step $i-1$, and the claim follows from the induction hypothesis. If $j\in\{j_L,j_R\}$ then $C=\{j_L,j_R\}$, and the claim follows. Otherwise, $j$ is strongly active at step $i$ and also at step $i-1$, and the claim follows from the induction hypothesis.
        \item $\vec{u}_i$ is of the form $(\deactivate,j_L,j_R)$. Then 
        \[V_i=V_{i-1}\text{ and } E_i=E_{i-1}\cup\{(f^{-1}(j_L,i),f^{-1}(j_R,i))\}\] 
        By the induction hypothesis, $S(j_L,i-1)=F(C_L)$ and $S(j_R,i-1)=f(C_R)$ where $C_L,C_R$ are the connected components of $f^{-1}(j_L,i-1),f^{-1}(j_R,i-1)$, respectively, in $G_{i-1}$. We now consider the following cases:
        \begin{itemize}
            \item $j\in (S(j_L,i-1)\cup S(j_L,i-1))\setminus\{j_L,j_R\}$, w.l.o.g. $j\in S(j_L,i-1)$. Since two vertices being connected by a path is an equivalence relation, we have $S(j_L,i-1)=S(j,i-1)$. In particular, by the induction hypothesis, $j$ is strongly active in step $i-1$ and the connected component of $f^{-1}(j,i-1)$ in $G_i$ is $C_L$. Therefore it is strongly active at step $i$ and the connected component of $f^{-1}(j,i)$ in $G_i$ is $C_L\cup C_R$. Additionally, by the definition of $S$, we have $S(j,i)=S(j_L,i-1)\cup S(j_R,i-1)=f(C_L)\cup f(C_R)=f(C_L\cup C_R)$, as needed.
            \item $j\in\{j_L,j_R\}$. Then $S(j,i)=\bot$ by definition and the claim follows.
            \item Otherwise, $j$ is either strongly active or inactive in both steps $i-1,i$, and there is no change in the connected component. Additionally, $S(j,i)=S(j,i-1)$, and the claim follows from the induction hypothesis.
        \end{itemize}
    \end{itemize}
    \noindent 
    Assume $G_\vec{u}$ has a cycle $C$. Every ``sweep'' is of degree at most $2$ in $G_\vec{u}$, as every sweep is activated at most once and deactivated at most once. In particular, all ``sweeps'' in $C$ have degree exactly $2$, and so $C$ is a connected component. Additionally, every such ``sweep'' is necessarily deactivated, or else it would have degree lower than $2$. Let $B_L,B_R\in P$ be the last vertices in $C$ that are deactivated, let $i$ be the index in which it happens and $j_L,j_R$ be the corresponding coordinates. Then $\vec{u}_i=(\deactivate,j_L,j_R)$, and $B_L,B_R$ are the only vertices $B\in C$ such that $f(B)$ are active at step $i$. Hence, they are also the only vertices $B\in C$ such that $f(B)$ is strongly active at step $i-1$, and by the claim we have $S(j_L,i-1)=S(j_R,i-1)=\{j_L,j_R\}$, as needed.
    \noindent 
    For the other direction, assume $\vec{u}_i=(\deactivate,j_L,j_R)$ and $S(j_L,i-1)=S(j_R,i-1)=\{j_L,j_R\}$. Then by the claim, $f^{-1}(j_L,i-1),f^{-1}(j_R,i-1)$ are connected in $G_{i-1}$. Additionally, $G_i$ is obtained from $G_{i-1}$ by adding an edge between them, forming a cycle.
\end{proof}
\cref{lem:sweep graph cycle characterization} essentially shows that an NFA can sequentially track an operation sequence and determine whether it corresponds to a run. 
It remains to deal with the requirement that $\sem{\vec{a}}_\Max\le k$.
\begin{defi}
\label{def:distance function}
    Let $\vec{u}$ be a $k$-operation sequence. The \emph{distance function of $\vec{u}$}, is the function $t\colon\{0,\ldots,n\}\times\{1,\ldots,\numslots\}\to2^{\{1,\ldots,\numslots\}}\cup\{\bot\}$ defined as follows. For $1\le j\le\numslots,1\le i\le n$:
    \begin{itemize}
        \item If $j$ is inactive at step $i$ then $t(j,i)=\bot$.
        \item Otherwise, $t(j,i)=i-i'$, where $i'$ is the maximal index such that $i'\le i$ and $\vec{u}_{i'}$ (either an $\activate$ or a $\none$ operation) involves $j$, or $t(j,i)=i$ if there does not exist such $i'$.
    \end{itemize}
\end{defi}
\noindent 
Intuitively, this captures the size of the jumps made by $\cA$. $t(j,i)$ is the number of steps that passed since we last read a letter into the relevant sweep.
Thus, we have that $\sem{\vec{a}}_\Max\le k$ if and only if the jumps prescribed by $t$ are bounded. More precisely, we have the following.
\begin{obs}
\label{obs:max semantics in terms of distance function}
    Let $k\in\bbN$ and $\vec{a}$ a jump sequence. Then $\sem{\vec{a}}_\Max\le k$ if and only if there exist a valid $k$-embedding $f$ 
    such that for the operation sequence $\vec{u}^{\vec{a},f}$ it holds that for every $1\le i\le n$ and $1\le j\le\numslots$ that is active at step $i$, we have $t(j,i)\le k-1$.
\end{obs}
We are now ready to describe the construction of our \NFA.
\shtodo{Got here.}
\begin{proof}[Proof of \cref{lem:max NFA construction}]
Denote $\cA=\tup{\Sigma,Q,\delta,Q_0,F}$. We construct $\cB=\tup{\Sigma,Q',\delta',Q'_0,F'}$ as follows:

$Q'=(\{R,L\}\times Q\times \{0,\ldots,k-1\}\times 2^{\{1,\ldots,\numslots\}})\cup\{\bot\})^\numslots$. Intuitively, $\cB$ guesses an operation sequence, which also dictates which coordinate every letter is read into, and simulates $\cA$ or its reverse on the chosen coordinate. After reading the $i$'th letter, each coordinate $j$ that is strongly active at step $i$ is marked by a tuple including the following information:
\begin{itemize}
    \item A flag specifying whether the sweep $f^{-1}(j,i)$ is a right or a left sweep.
    \item The current state in the simulated run of $\cA$ on $w_{f^{-1}(j,i)}$, where $w_{f^{-1}(j,i)}$ is the turning subword $w_\vec{b}$ for the sweep $\vec{b}$ given by $f^{-1}(j,i)$. 
    \item $t(j,i)$ (where $t$ is the distance function).
    \item $S(j,i)$ (where $S$ is the component function).
\end{itemize}
If $j$ is inactive or weekly active at step $i$, it is marked by $\bot$.

$Q'_0=\{(R,q_0,0,\{1\})\mid q_0\in Q_0\}\times\{\bot\}^{\numslots-1}$.

\noindent 
$F'=\{(\vec{v}_1,\ldots\vec{v}_\numslots)\mid$ there exists $j$ such that $\vec{v}_j=(R,q_f,t,\{j\})$, where $q_f\in F$, and $\vec{v}_{j'}=\bot$ for all $j'\neq j\}$.

We proceed to define $\delta'$ as a case split of three cases,
corresponding to the different possible operations. A vector $(\vec{v}'_1,\ldots,\vec{v}'_\numslots)$ is in $\delta'((\vec{v}_1,\ldots,\vec{v}_\numslots),\sigma)$ if it satisfies one of the following conditions:
\begin{itemize}
    \item Transition corresponding to $\none$ operations: There exists $1\le j\le \numslots$ such that:
    \begin{itemize}
        \item Either $\vec{v}_j=(R,q_j,t_j,S_j),\vec{v'}_j=(R,q'_j,0,S_j)$ where $q'_j\in\delta(q_j,\sigma)$, or $\vec{v}_j=(L,q_j,t_j,S_j),$ $\vec{v'}_j=(L,q'_j,0,S_j)$ where $q_j\in\delta(q'_j,\sigma)$.
        \item For all $j'\neq j$, either $\vec{v}_{j'}=\vec{v}'_{j'}=\bot$, or $\vec{v}_{j'
        }=D_{j'},q_{j'},t_{j'},S_{j'}$ and $\vec{v}'_{j'
        }=D_{j'},q_{j'},t_{j'}+1,S_{j'}$
        Note that $t_{j'}+1$ must not exceed $k$, reflecting the fact that we cannot go beyond jumping budget $k$.
    \end{itemize}
    \item Transition corresponding to $\activate$ operations: There exist $1\le l_L,j_R\le\numslots$ such that:
    \begin{itemize}
        \item $\vec{v}_{j_L}=\vec{v}_{j_R}=\bot$.
        \item $\vec{v}'_{j_R}=(R,q,0,\{j_L,j'_R\})$.
        \item $\vec{v}'_{j_L}=(L,p,0,\{j_L,j_R\})$ such that $q\in\delta(p,\sigma)$.
        Intuitively, $q$ is the ``turning state'' between the two sweeps, and since a letter is read into coordinate $j_L$, its state is set to $p$.
        \item For all $j\notin \{j_L,j_R\}$, either $\vec{v}_lj=\vec{v}'j=\bot$, or $\vec{v}_j=(j,q_j,t_j,S_j)$ and $\vec{v}'_jl=(D_lj,q_lj,t_j+1,S_j)$.
    \end{itemize}
    \item Transition corresponding to $\deactivate$ operations: There exist $1\le j_L,j_R\le\numslots$ such that:
    \begin{itemize}
        \item $\vec{v}_{j_L}=(L,q',t_{j_L},S_{j_L})$.
        \item $\vec{v}_{j_R}=(R,q,t_{j_R},S_{j_R})$, where $q'\in\delta(q,\sigma)$.
        \item $\vec{v'}_{j_L}=\vec{v'}_{j_R}=\bot$.
        \item For all $j\notin \{j_L,j_R\}$, either $\vec{v}_j=\vec{v}'_j=\bot$, or $\vec{v}_j=(D_j,q_j,t_j,S_j)$ and $\vec{v}'_j=(D_j,q_j,t_j+1,S'_j)$, such that if $j\notin S_{j_L}\cup S_{j_R}$ then $S'_j=S_j$, and if $j\in S_{j_L}\cup S_{j_R}$ then $S'_j=S_{j_L}\cup S_{j_R}$.
        \item It \emph{does not} hold that $S_{j_L}=S_{j_R}=\{j_L,j_R\}$.
    \end{itemize}
\end{itemize}
\noindent 
It remains to prove the correctness of the construction.
Assume $\cA_\Max(w)\le k$, with a corresponding jump sequence $\vec{a}$. By \cref{lem:max number of simultaneous sweeps}, $\vec{a}$ has a valid $k$-embedding $f$. Let $\vec{u}=\vec{u}^{\vec{a},f}$. By \cref{lem:G is a path <-> there exists a jump sequence}, $G_\vec{u}$ is a path, whose first vertex $B$ must have $f(B)=1$ and therefore $\vec{u}$ is $1$-valid. Consider the run $\rho$ of $\cB$ that ``guesses the operations'' of $\vec{u}$ and correctly simulates accepting runs of $\cA$ or its reverse on each of the sweeps. Such a run exists with the constraint on the distance function satisfied by \cref{obs:max semantics in terms of distance function}, and the condition $S_{j_L}=S_{j_R}=\{j_L,j_R\}$ never occurs upon an operation $(\deactivate,j_L,j_R)$ by \cref{lem:G is a path <-> there exists a jump sequence,lem:sweep graph cycle characterization}. Hence $\cB$ accepts $w$.

Conversely, assume $w\in\regL(\cB)$ with an accepting run $\rho$. Let $\vec{u}$ be the operation sequence guessed by $\rho$, and $P,f$ be the partition of $\{1,\ldots,n\}$ and embedding corresponding to $\vec{u}$. Every vertex in $G_\vec{u}$ has degree at most $2$. $G_\vec{u}$ is the disjoint union of either a simple path (if $\vec{u}$ is $1$-valid) or an isolated vertex (otherwise), and any number of cycles. By \cref{lem:sweep graph cycle characterization}, $G_\vec{u}$ has no cycles, and is therefore a simple path, and in particular, $\vec{u}$ is $1$-valid. By \cref{lem:G is a path <-> there exists a jump sequence}, There exists a jump sequence $\vec{a}$ such that $\vec{u}=\vec{u}^{\vec{a},f}$. Since $\rho$ is accepting, we have that $w_\vec{a}\in\regL(\cA)$. Additionally, by \cref{obs:max semantics in terms of distance function} we have $\sem{\vec{a}}_\Max\le k$, and so $\cA_\Max(w)\le k$.
\end{proof}

We can now conclude the decidability of boundedness in the usual manner.
\begin{thm}
\label{thm:max decidable boundedness}
The following problems are decidable for the $\Max$ semantics: \probDomkBoundedness,\  \probDomParamBoundedness,\ \probkBoundedness, and \probParamBoundedness.
\end{thm}

The proof of~\cref{lem:max NFA construction} shows that the size of $\cB$ is polynomial in the size of $\cA$ and single-exponential in $k$, giving us $\cxPSPACE$ membership for \probkBoundedness, and $\cxEXPSPACE/$ $\cxEXPEXPSPACE$ for \probParamBoundedness with unary/binary encoding, respectively.

Note that as with the other semantics, we trivially get PSPACE-hardness for \probUnivBoundedness{0} (and therefore hardness for \probParamBoundedness and \probDomParamBoundedness). For the case of \probkBoundedness for $k\neq 0$, however, we leave the lower bound open. We note that the hardness techniques used for the other semantics fail for the $\Max$ semantics. Specifically, this is due to the fact that unlike the other semantics, $\Max$ is not cumulative: that is, once a prefix of a jump sequence attains cost $k$, even if the rest of the permutation has additional cost, the overall cost might still be $k$. This means that we cannot use the $\fsh$-prefixes as in other hardness proofs to ``max out'' the jumping budget.

\section{Interplay Between the Semantics}
\label{sec:interplay}
Having established some decidability results, we now turn our attention to the interplay between the different semantics, in the context of boundedness. 
We show that for a given \JFA $\cA$, if $\cA_\Abs$ is bounded, then so are $\cA_\Max$ and $\cA_\Ham$, if $\cA_\Ham$ is bounded, then so is $\cA_\Rev$, and if $\cA_\Max$ is bounded, then so is $\cA_\Rev$. 
We complete the picture by showing that these are the only relationships -- we give examples for the remaining cases (see~\cref{tab:separation}). 

\begin{lem}
\label{lem:relationship Abs to Max}
Consider a $\JFA$ $\cA$. If $\cA_\Abs$ is bounded, then  $\cA_\Max$ is bounded.
\end{lem}
\begin{proof}
    For every jump sequence $\vec{a}$ we have $\sem{\vec{a}}_\Max\le\sem{\vec{a}}$. It follows that for all $k\in\bbN,w\in\Sigma^*$, if there exists a jump sequence $\vec{a}$ such that $w_\vec{a}\in\regL(\cA)$ and $\sem{\vec{a}}\le k$, then we also have $\sem{\vec{a}}_\Max\le k$, and the claim follows.
\end{proof}

\begin{lem}
	\label{lem:relationship Abs to Ham}
	Consider a $\JFA$ $\cA$. If $\cA_\Abs$ is bounded, then  $\cA_\Ham$ is bounded.
\end{lem}
\begin{proof}
	Consider a word $w\in \Sigma^*$, we show that if $\cA_\Abs(w)\le k$ for some $k\in \bbN$ then $\cA_{\Ham}(w)\le (2k+1)(k+1)$. 
	Assume $\cA_\Abs(w)\le k$, then there exists a jump sequence $\vec{a}=(a_0,\ldots,a_{n+1})$ such that $\sem{\vec{a}}\le k$ and $w_\vec{a}\in \regL(\cA)$.
	In the following we show that $a_i=i$ for all but $(2k+1)(k+1)$ indices, i.e.,
	 $|\{i\mid a_i\neq i\}|\le (2k+1)(k+1)$. 

  It is convenient to think of the jumping head moving according to $\vec{a}$ in tandem with a sequential head moving from left to right. 
  Recall that by~\cref{lem:abs window}, for every index $i$ we have that $i-k\le a_i\le i+k$, i.e. the jumping head stays within distance $k$ from the sequential head. 
	 
	 Consider an index $i$ such that $a_i\neq i$ (if there is no such index, we are done).
	 We claim that within at most $2k$ steps, $\cA$ performs a jump of cost at least 1 according to $\vec{a}$. More precisely, there exists $i+1\le j\le i+2k$ such that $|a_{j}-a_{j-1}|>1$. To show this we split to two cases:
	 \begin{itemize}
	 	\item If $a_i>i$, then there exists some $m\le i$  such that $m$ has not yet been visited according to $\vec{a}$ (i.e., by step $i$). Index $m$ must be visited by $a_i$ within at most $k$ steps (otherwise it becomes outside the $i-k,i+k$ window around the sequential head), and since $a_i>i$, it must perform a `` left jump'' of size at least 2 (otherwise it always remains to the right of the sequential reading head). 
	 	\item If $a_i<i$, then there exists some $m\ge i$ such that $m$ has already been visited by step $i$ according to $\vec{a}$. Therefore, within at most $2k$ steps, the jumping head must skip at least over this position (think of $m$ as a hurdle coming toward the jumping head, which must stay within distance $k$ of the sequential head and therefor has to skip over it). Such a jump incurs a cost of at least 1.
	 \end{itemize}
	Now, let $B=\{i\mid a_i\neq i\}$ and assume by way of contradiction that $|B|> (2k+1)(k+1)$. By the above, for every $i\in B$, within $2k$ steps the run incurs a cost of at least 1. While some of these intervals of $2k$ steps may overlap, we can still find at least $k+1$ such disjoint segments (indeed, every $i\in B$ can cause an overlap with at most $2k$ other indices). More precisely, there are $i_1<i_2<\ldots<i_{k+1}$ in $B$ such that $i_j>i_{j-1}+2k$ for all $j$, and therefore each of the costs incurred within $2k$ steps of visiting $i_j$ is independent of the others. 
 This, however, implies that $\sem{\vec{a}}\ge k+1$, which is a contradiction, so $|B|\le (2k+1)(k+1)$.
	
	It now follows that $\cA_\Ham(w)=|\{i\mid w_{a_i}\neq w_i\}|\le |\{i\mid a_i\neq i\}|\le (2k+1)(k+1)$  	
\end{proof}
\begin{lem}
    \label{lem:relationship Ham to Rev}
    Consider a $\JFA$ $\cA$. If $\cA_\Ham$ is bounded, then  $\cA_\Rev$ is bounded.    
\end{lem}
\begin{proof}
	Consider a word $w\in \Sigma^*$, we show that if $\cA_\Ham(w)\le k$ for some $k\in \bbN$ then $\cA_{\Rev}(w)\le 3k$. 
	Assume $\cA_\Ham(w)\le k$, then there exists a jump sequence $\vec{a}=(a_0,\ldots,a_{n+1})$ such that $w_\vec{a}\in \regL(\cA)$ and $w_\vec{a}$ differs from $w$ in at most $k$ indices. 
	We claim that we can assume without loss of generality that for every index $i$ such that $w_{a_i}= w_i$ we have $a_i=i$ (i.e., $i$ is a \emph{fixed point}). 
	Intuitively -- there is no point swapping identical letters.
	Indeed, assume that this is not the case, and further assume that $\vec{a}$ has the minimal number of fixed-points among such jump sequences. Thus, there exists some $j$ for which ${a_j}\neq j$ but $w_{a_j}=w_j$. Let $m$ be such that $a_m=j$, and consider the jump sequence $\vec{a'}=(a'_0,\ldots,a'_{n+1})$ obtained from $\vec{a}$ by composing it with the swap $(a_j\  a_m)$. 
	Then, for every $i\notin \{j,m\}$ we have that $a'_i=a_i$. In addition, $a'_j=a_m=j$ as well as $a'_m=a_j$. In particular, $\vec{a'}$ has more fixed points than $a$ (exactly those of $\vec{a}$ and $j$). However, we claim that $w_\vec{a}=w_\vec{a'}$. Indeed, the only potentially-problematic coordinates are $a_j$ and $a_m$. For $j$ we have $w_{a_j}=w_j=w_{a'_j}$. 
	and for $m$ we have $w_{a'_m}=w_{a_j}=w_j=w_{a_m}$.
	This is a contradiction to $\vec{a}$ having a minimal number of fixed points, so we conclude that no such coordinate $a_j\neq j$ exists.
	
	Next, observe that $\turnindices(\vec{a})\subseteq \{i\mid a_i\neq i\vee a_{i+1}\neq i+1 \vee a_{i-1}\neq i-1\}$. Indeed, if $a_{i-1}=i-1$, $a_i=i$ and $a_{i+1}=i+1$ then clearly $i$ is not a turning index. By the property established above, we have that $w_{a_i}=w_i$, if and only if $a_i=i$. It follows that $\turnindices(\vec{a})\subseteq \{i\mid w_{a_i}\neq w_{i}\vee w_{a_{i+1}}\neq w_{i+1} \vee w_{a_{i-1}}\neq w_{i-1}\}$, so $|\turnindices(\vec{a})|\le 3k$ (since each index where $w_\vec{a}\neq w$ is counted at most 3 times\footnote{A slightly finer analysis shows that this is in fact at most  $2k$, but we are only concerned with boundedness.} in the latter set). 
\end{proof}
Combining~\cref{lem:relationship Ham to Rev,lem:relationship Abs to Ham}, we have the following.
\begin{cor}
	\label{cor:relationship Abs to Rev}
    Consider a $\JFA$ $\cA$. If $\cA_\Abs$ is bounded, then $\cA_\Rev$ is bounded.
\end{cor}

The $\Max$ semantics poses a greater challenge than other semantics. Nonetheless, we are able to relate it to the other semantics.
\begin{lem}
    \label{lem:relationship Max to Rev}
    Consider a $\JFA$ $\cA$. If $\cA_\Max$ is bounded, then  $\cA_\Rev$ is bounded.    
\end{lem}
Before proving the lemma, we first demonstrate the difficulty in proving it. Observe that for each of the previous implications between the semantics regarding boundedness, in order to show that a word $w$ with low cost in one semantics also has low cost in another, we actually show that a jumping sequence $\vec{a}$ that ``witnesses'' this low cost in the one semantics, also witnesses it in the other.
For example,
in the proof of \cref{lem:relationship Abs to Ham}, given $w$ and $\vec{a}$ such that $\sem{\vec{a}}\le k$, we showed that $d_H(w,w_\vec{a})\le (2k+1)(k+1)$, and similar results hold in the contexts of \cref{lem:relationship Abs to Max,lem:relationship Ham to Rev}. 

Unfortunately, an analogous claim does not hold for the relationship between $\Max$ and $\Rev$. Indeed, for $w,\vec{a}$ such that $\sem{\vec{a}}_\Max\le k$ the value $\min\{\numrev(\vec{b})\mid w_\vec{b}=w_\vec{a}\}$ can be unboundedly large. 
Stated precisely, the following example demonstrates the existence of a (non-regular) language $L$ and a language $L'$ such that every $w\in L'$ has a permutation in $L$, with $\{\min\{\sem{\vec{a}}_\Max\mid w_\vec{a}\in L\} \mid w\in L'\}$ being bounded, but $\{\min\{\numrev(\vec{a})\mid w_\vec{a}\in L\} \mid w\in L'\}$ is unbounded.
\begin{exa}
    \label{xmp:bounded max unbounded rev}
    For $n\in\bbN$, let $w^{(n)}=abcaabca^4bc\cdots a^{2^n}bc=(a^{2^i}bc)_{i=0}^n$, and $w'^{(n)}=acbaacba^4cb\cdots a^{2^n}cb=(a^{2^i}cb)_{i=0}^n$. 
    Consider $L=\{w^{(n)}\mid n\in\bbN\},L'=\{w'^{(n)}\mid n\in\bbN\}$. Observe that, while $L$ is unbounded in the $\Max$ semantics, we do get bounded $\Max$ when restricting the input to $L'$. That is, every $w'^{(n)}\in L'$ has a permutation $w^{(n)}\in L$ induced by the jump sequence $\vec{a}$ which ``swaps every $cb$ into $bc$'', which yields $\sem{\vec{a}}_\Max=1$ and $w'^{(n)}_\vec{a}=w^{(n)}$. 
    However, we show that $\min\{\numrev(\vec{b})\mid w'^{(n)}_\vec{b}=w^{(n)}\}$ is unboundedly large as $n$ increases. 
    Since $w^{(n)}$ is the only permutation of $w'^{(n)}$ belonging to $L$, the claim follows.

    Before proving this formally, we give an intuitive approach. Consider the words
    \[
    \begin{split}
        &w^{(n)}=abcaabcaaaabca^8bca^{16}bca^{32}bca^{64}bc\\
        &w'^{(n)}=acbaacbaaaacba^8cba^{16}cba^{32}cba^{64}cb\\
    \end{split}
    \]
    and try to construct a jumping run on $w'^{(n)}$ that reads $w^{(n)}$, with very few reversals. In order to minimize reversals, it seems reasonable to skip the first $c$, and then obtain the $bc$ pairs by taking $b$ from one pair with the $c$ from the next $cb$. 
    However, in order to read enough $a$'s in between pairs, one must skip some $b$'s as well.
    Then, after one sweep, we somehow need to go back and pick the remaining $b$'s and $c$'s, together with enough $a$'s in between. There, problems start to arise -- if we leave too many ``holes'' in the word, we find ourselves needing to go back and forth to pick up $a$'s.
    This is far from any formal argument, but playing around with this approach gives a strong feeling that the reversals should be unbounded. 
    We now turn to give a formal argument for this.
    
    Assume by way of contradiction that there exists $k$ such that $\min\{\numrev(\vec{b})\mid w'^{(n)}_\vec{b}=w^{(n)}\}\le k$ for all $n$. Consider $n=100k^2$ and $w'=w'^{(n)}$, and let $\vec{b}$ be a jump sequence such that $\numrev(\vec{b})\le k$ and $w'_\vec{b}=w^{(n)}$. We partition $\{1,\ldots,|w'|\}$ into $k$ segments, each corresponding to a sub-word of $w'$ consisting of $100k$ consecutive sequences of the form $a^*cb$. That is, for every $i\in\{1,\ldots, k\}$, let $n_i=\sum_{j=100k(i-1)+1}^{100ki}(2^j+2)$ (note that each $2^j+2$ is the length of the $j$-th $a^*cb$ sequence), and the $i$'th segment is then $S_i=\{\sum_{j=1}^{(i-1)}n_j+1,\ldots,\sum_{j=1}^in_j\}$. 
    In addition, let 
    \[B^{w'}_i=\{j\in S_i\mid w'_j=b\} \quad B^\vec{b}_i=\{l\in\{1,\ldots, |w'|\}\mid w'_l=b\wedge\exists j\in S_i:\vec{b}_j=l\}\] 
    That is, we consider the set $B^{w'}_i$ of occurrences of the letter $b$ that belong to the $i$'th segment of $w'$, versus the set of occurrences of $b$ that are read by the $i$'th segment of $\vec{b}$.  
    As a sanity check, if $\vec{b}$ is the jump sequence that always flips the pairs $cb$ to $bc$ (which is not bounded-reversal), namely 
    \[
    \vec{b}=(0,1,3,2,4,5,7,6,8,9,10,11,13,12,\ldots)
    \]
    then we would have $B_i^{w'}=B_i^{\vec{b}}$, since all $b$'s are read in their original segment. Specifically, we would have (depending on $k$) e.g., $B_1^{w'}=B_1^{\vec{b}}=\{3,7,13,\ldots\}$, where the ``reason'' for $3\in B_1^{w'}$ is because $\vec{b}_2=3$. 
    Another way to view $B^{\vec{b}}_i$ is as the set of occurrences of $b$ that are read at \emph{times} that correspond the $i$-th segment.

   Fix $i\in\{1,\ldots,k\}$ and let $A=(\bigcup_{j=1}^iB^{w'}_j)\setminus(\bigcup_{j=1}^iB^\vec{b}_j)$ 
   be the set of occurrences of $b$ that have not yet been read by $w'_{\vec{b}}$ after reading segments $1,\ldots, i$.
   We claim that $|A|\le 2k$ (actually, $|A|\le k+1\le 2k$, but $2k$ is more convenient to work with). 
   That is, after reading $i$ segments, at most $2k$ of the $b$'s belonging to the first $i$ segments of $w'$ remain unread. Intuitively, if there are many $b$'s appearing at the beginning of $w'$ that are read at a late stage of the jump sequence, then between every two such $b$'s we must read a huge number of $a$'s -- more $a$'s than those available around the first $i$ segments, and thus we must make a jump to the right and back to the left in order to read sufficiently many $a$'s, resulting in too many reversals. 
   
   Formally, consider the set $C=\{m\mid \vec{b}_m\in A\}$, i.e., the inverse image under $\vec{b}$ of the set $A$. 
   Intuitively, $m\in C$ means that at timestep $m$ we have $w'_{\vec{b}_m}=b$, but $\vec{b}_m$ is before the end of $S_i$, whereas $m$ is after. 
   We order the elements of $C$ as $m_1<\ldots <M_{|A|}$.

   Intuitively, we claim that once $b$ is read at time $m_l$ for some $1\le l<|A|$, then before reading $b$ again at $m_{l+1}$, so many $a$'s need to be read, that a segment above $i$ must be visited, which is only possible with a reversal.

   More precisely, for every $l<|A|$ let $D_l=\{t\mid m_l<t<m_{l+1}\wedge w_t=a\}$. Since $m_l\in\bigcup_{j>i}S_j$ we have that $|D_l|\ge2^{100ki+1}$ (since we need to read at least this many $a$'s between consecutive $b$ at such late times). 
   For all $t\in D_l$ we have that $w'_{\vec{b}_t}=a$, but $|\{t\in \bigcup_{j=1}^i S_j\mid w'_t=a\}|=\sum_{r=1}^{100ki}2^r<2^{100ki+1}$, and so there exists $t\in D_l\setminus\bigcup_{j=1}^i S_j$. 
      
   Since $\vec{b}_{m_l},\vec{b}_{m_{l+1}}\in A\subseteq\bigcup_{j=1}^i S_j$ we conclude that $\vec{b}$ performs at least one reversal between the indices $m_l,m_{l+1}$. 
   Thus, $\vec{b}$ performs at least $|A|-1$ reversals overall, and so $|A|\le\numrev(\vec{b})+1\le k+1\le 2k$.

   We now claim that, intuitively, the $b$'s of $w'$ are read ``mostly sequentially'' in the following sense: Denote by $B_i=B^{w'}_i\cap B^\vec{b}_i$ the set of ``well-behaved'' $b$'s -- those that are in segment $i$ and are read at a time corresponding to segment $i$. 
   We claim that $|B_i|\ge 96k$ for all $i$. 
   Indeed, $|B^{w'}_i|=100k$ by definition, and observe that any occurrence of $b\in B_i^{w'}$ that is read ``too early'' (i.e., read in segment $j<i$) is by definition not itself in segment $j<i$ (this is a trivial observation), i.e., 
   \[B^{w'}_i\cap\left(\bigcup_{j=1}^{i-1}B^\vec{b}_j\right)\subseteq\left(\bigcup_{j=1}^{i-1}B^\vec{b}_j\right)\setminus\left(\bigcup_{j=1}^{i-1}B^{w'}_j\right)\]
   Additionally, we observe that there are equal numbers of $b$'s that are read in \emph{times} before segment $i$, but are not in locations before segment $i$, as there are $b$'s that are in locations before segment $i$, but are not read in times before segment $i$. That is,
   \[\left|\left(\bigcup_{j=1}^{i-1}B^\vec{b}_j\right)\setminus\left(\bigcup_{j=1}^{i-1}B^{w'}_j\right)\right|=\left|\left(\bigcup_{j=1}^{i-1}B^{w'
   }_j\right)\setminus\left(\bigcup_{j=1}^{i-1}B^\vec{b}_j\right)\right|\] 
   Indeed, this follows since $|\bigcup_{j=1}^{i-1}B^{w'}_j|=|\bigcup_{j=1}^{i-1}B^\vec{b}_j|$, and in both sides we remove the same set of ``well-behaved'' $b$'s.
   
   Thus, by the bound on $|A|$ and since $B^{w'}_i\cap(\bigcup_{j=1}^{i-1}B^\vec{b}_j)\subseteq A$, we have $|B^{w'}_i\cap(\bigcup_{j=1}^{i-1}B^\vec{b}_j)|\le 2k$. 
   Since any $b$ that is after segment $i$ and read later than segment $i$, was in particular not read up to segment $i$, we have
   \[ B^{w'}_i\cap\left(\bigcup_{j>i}B^\vec{b}_j\right)\subseteq\left(\bigcup_{j=1}^iB^{w'}_j\right)\setminus\left(\bigcup_{j=1}^iB^\vec{b}_j\right)\subseteq A
   \]  
   So again by the bound on $|A|$ we now have
   \[
   \begin{split}
   &|B_i|\ge |B_i^{w'}\cap B_i^{\vec{b}}|=\left|B_i^{w'}\setminus  \left(\bigcup_{j=1}^{i-1}B_j^{\vec{b}}\cup \bigcup_{j>i}B_j^{\vec{b}}\right)\right|\ge\\
   &|B_i^{w'}|-\left|B_i^{w'}\cap \bigcup_{j=1}^{i-1}B_j^{\vec{b}}\right|-\left|B_i^{w'}\cap\bigcup_{j>i}B_j^{\vec{b}}\right|=100k-2k-2k=96k
   \end{split}
   \]

   We next claim that for all $i$, $B_i$ is not read monotonically, that is, there exist $j^i_1<j^i_2\in S_i$ such that $\vec{b}_{j^i_2}<\vec{b}_{j^i_1}\in B_i$. Intuitively, it is impossible to read the majority of $b$'s of a segment from left to right without making a reversal between many pairs of subsequent $b$'s, resulting in too many reversals. Indeed, assume by way of contradiction that is not the case, and so there exist $m_1<\ldots<m_{|B_i|}\in S_i$ such that $\vec{b}_{m_1}<\ldots<\vec{b}_{m_{|B_i|}}\in B_i$. We have $|B^{w'}_i\setminus B_i|\le 4k$, and so denoting $B^{w'}_i=\{l_1<\ldots<l_{100k}\}$, there exist at most $8k$ indices of the form $l_t\in B^{w'}_i$ such that at least one of $l_t,l_{t+1}$ is not in $B_i$. Hence, there exist at least $92k$ indices $l_t\in B_i^{w'}$ such that $l_t,l_{t+1}\in B_i$. For each such index $l_t$, consider the infix of $w'$ given by
   $w'_{\vec{b}_{l_t}},w'_{\vec{b}_{l_t}+1},\ldots, w'_{\vec{b}_{l_t+1}-1},w'_{\vec{b}_{l_{t+1}}}$. Note that this is a contiguous infix of $w'$ from index $\vec{b}_{l_t}$ of length $l_{t+1}-l_t$. By our assumption (that the $b$'s are read consecutively), it follows that this infix is of the form $ba^*cb$, as a consecutive infix of $w'$ between two occurrences of $b$.

   We now compare this infix to what is actually read following $\vec{b}$. While reading $b$'s happens consecutively within $B_i^{w'}$, outside it the behavior can be arbitrary, with the exception that the resulting word is $w'$. 
   We therefore have that, $w'_{\vec{b}_{l_t}}w'_{\vec{b}_{l_t+1}}\cdots w'_{\vec{b}_{l_{t+1}-1}}\ w'_{\vec{b}_{l_{t+1}}}\in (bca^*)^*b$. 
   In particular,  between reading the two $b$'s we must at some point read $ca$. However, $ca$ is not available in the consecutive infix above. Therefore, there must be at least one reversal in $\vec{b}$ during this infix. 
   Since this is true in every one of the $92k$ indices we consider,    
   it follows that $\numrev(\vec{b})\ge 92k>k$, thus leading to a contradiction.

   To finalize the proof, this contradiction means that most $b$'s of each segment are read by the corresponding segment of the jump sequence, thus incurring an overall ``nearly sequential'' run on the $b$'s, but each of the $k$ segments also incurs a reversal within the segment. Since we have $k$ segments, this sums up to at least $2k$ reversals (i.e., two turning points per reversal). More precisely, for all $i$ we have $S_i<S_{i+1}$ and $B_i<B_{i+1}$, and so $0<j^1_1<j^1_2<j^2_1<j^2_2<\ldots<j^k_1<j^k_2<n+1$, while $\vec{b}_0<\vec{b}_{j^1_2}<\vec{b}_{j^1_1}<\vec{b}_{j^2_2}<\vec{b}_{j^2_1}<\ldots<\vec{b}_{j^k_2}<\vec{b}_{j^k_1}<\vec{b}_{n+1}$, resulting in $\numrev(\vec{b})\ge2k>k$, leading to a contradiction.
\end{exa}
The example shows that in proving \cref{lem:relationship Max to Rev}, there is no hope to take a jump sequence $\vec{a}$ with bounded $\sem{\vec{a}}_\Max$ and show that it induces bounded reversals, nor that we can modify it and obtain the same word using another jump sequence with bounded reversals. We must actually find an entirely different word that is accepted with a different jump sequence with bounded reversals, but is still in the target language.

We proceed with the proof of \cref{lem:relationship Max to Rev}. Intuitively, 
we show that the jump sequences witnessing the bounded $\Max$ induce words consisting of boundedly many sequences of the same letter (e.g., $a^*b^*c^*$). These words are in $\regL(\cA)$ by definition, and in turn, they witness a bounded $\Rev$ cost for every $w\in\jL(A)$.

\begin{proof}[Proof of \cref{lem:relationship Max to Rev}]
    Let $\cA$ be a JFA such that $\cA_\Max(w)\le k$ for all $w\in\jL(\cA)$. Let $w\in\jL(\cA)$. Write $\Sigma=\{\sigma_1,\ldots,\sigma_{|\Sigma|}\}$, and for all $i\in\{1,\ldots, |\Sigma|\}$, let $m_i$ be the number of $\sigma_i$ occurrences in $w$. Then $w'=\sigma_1^{m_1}\cdots\sigma_{|\Sigma|}^{m_{|\Sigma|}}\in\jL(\cA)$, and therefore has a jump sequence $\vec{a}$ such that $\sem{\vec{a}}_\Max\le k$ and $w'_\vec{a}\in\regL(\cA)$. By \cref{lem:bounded max => bounded cut-crossing} we have that $\numcross(\vec{a},\sum_{j=1}^im_j)\le 2k+1$ for all $1\le i\le|\Sigma|-1$. It follows that $w'_\vec{a}\in\{\sigma_1^*+\ldots+\sigma_{|\Sigma|}^*\}^{k'}$ for $k'=(|\Sigma|-1)(2k+1)+1$, since every transition between sequences of two different letters crosses at least one of the $|\Sigma|-1$ corresponding cuts. Observe that $w'_\vec{a}$ is also a permutation of $w$, and we claim that there exists a jump sequence $\vec{b}$ such that $w_\vec{b}=w'_\vec{a}$ and $\numrev(\vec{b})\le k'$. Indeed, writing $w'_\vec{a}=\sigma_{l_1}^{r_1}\cdots\sigma_{l_{k'}}^{r_{k'}}$, the above can be achieved by reading any $r_1$ indices of $w$ containing $\sigma_{l_1}$ from left to right, then $r_2$ occurrences of $\sigma_{l_2}$ from right to left, and so on. We therefore have that $\cA_\Rev(w)\le k'$ for all $w\in\jL(\cA)$.
\end{proof}

We proceed to show that no other implication holds with regard to boundedness, by demonstrating languages for each possible choice of bounded/unbounded semantics (c.f.~\cref{rmk:semantics semantics}). The examples are summarized in~\cref{tab:separation}, and are proved below.
        

\begin{table}[ht]
    \centering
    \begin{tabular}{ccccc}
        \toprule
        $\Abs$ & $\Ham$ & $\Rev$ & $\Max$ & Language \\
        \midrule
        Bounded & Bounded & Bounded & Bounded & $(a+b)^*$ \\
        Unbounded & Bounded & Bounded & Bounded & $c^*ac^*bc^*$ \\
        Unbounded & Bounded & Bounded & Unbounded & $(a+b)^*a$ \\
        Unbounded & Unbounded & Bounded & Bounded &
            \makecell{$((a(aa)^*b(bb)^*)^*+$ \\ $(b(bb)^*a(aa)^*)^*)(a^*+b^*)$} \\
        Unbounded & Unbounded & Bounded & Unbounded & $a^*b^*$ \\
        Unbounded & Unbounded & Unbounded & Unbounded & $(ab)^*$ \\
        \bottomrule
    \end{tabular}
    \caption{Examples for every possible combination of bounded/unbounded semantics. The languages are given by regular expressions (e.g., $(a+b)^*a$ is the language of words that end with $a$.)}
    \label{tab:separation}
\end{table}

\begin{exa}
	\label{xmp:BBBB}
	The language $(a+b)^*$ is bounded in all semantics. This is trivial, since every word is accepted, and in particular has cost $0$ in all semantics.
\end{exa}

\begin{exa}
    \label{xmp:UBBB}
    The language $c^*ac^*bc^*$ is unbounded in the $\Abs$ semantics, but bounded in $\Ham$, $\Rev$ and $\Max$.
    Indeed, let $\cA$ be an \NFA such that $\regL(\cA)=c^*ac^*bc^*$ and consider a word $w\in \jL(\cA)$. If $w\in\regL(\cA)$, then trivially $\cA_\Ham(w)=0$. Otherwise, $w\in c^*bc^*ac^*$, and $\cA_\Ham(w)=2$. Overall, $\Ham$ is bounded, and by \cref{lem:relationship Ham to Rev}, so is $\Rev$. 
    
    For $\Max$, we again only need to consider words $w\in c^*bc^*ac^*$. Intuitively, in order to read $a$ before $b$, we must jump over $b$, but leave enough $c$'s unread so that we can go back from $a$ to $b$, and then forward again until the end of the word.
    Consider the case that $w\in c^*b(ccc)^*ac^*$, that is, there exist $i,j$ such that $w_i=b$, $w_{i+3j+1}=a$, and $w_k=c$ for all other $k$. Consider the jump sequence $\vec{a}=(0,1,\ldots,i-1,i+1,i+4,i+7,\ldots i+3j+1,i+3j,i+3j-3,\ldots i,i+2,i+5,\ldots i+3j-1,i+3j+2,i+3j+3,\ldots n+1)$. That is, we first read the first sequence of $c$. We then jump over the $b$ and read every third $c$ until reaching $a$. After reading the $a$, we turn and read every third $c$ while moving left until reaching $b$ and reading it, then turn again and read the remaining $c$'s in increasing order. We have $\sem{\vec{a}}_\Max=2$. The analysis for $w\in c^*b(ccc)^*cac^*$ and $w\in c^*b(ccc)^*ccac^*$ is similar, and we have that $\Max$ is bounded.

    For $\Abs$, however, consider the word $w=bc^na$ for every $n\in \bbN$. Any accepting jump sequence $\vec{a}$ has $\vec{a}_j=n+2,\vec{a}_k=1$ for $j<k$. Let $A\subseteq\{2,\ldots,n+1\}=\{\vec{a}_i\mid i<j\}$. That is, $A$ corresponds to all those $c$'s that are read before the $a$. Denote $A=\{i_1,\ldots,i_{|A|}\}$ and $\overline{A}=\{i_{|A|+1},\ldots,i_n\}$. Now,
    \begin{align*}
        \sem{\vec{a}} = & \sum_{i=1}^{n+1}\sem{a_i-a_{i-1}}\ge \\
        = & \sum_{i=1}^{j}\sem{a_i-a_{i-1}} + \sum_{i=j+1}^{k}\sem{a_i-a_{i-1}} \\
        \overset{(*)}{\ge} & (|A|+1) + (n-|A|) \\
        = & n+1
    \end{align*}
    Where $(*)$ is due to the fact that every "skipped" letter adds $1$ to the size of at least one jump. By increasing $n$, we have that $\Abs$ is unbounded.
\end{exa}

\begin{exa}
    \label{xmp:UBBU}
    The language $(a+b)^*a$ is bounded in the $\Ham$ and $\Rev$ semantics, but unbounded in $\Abs$ and $\Max$. 
    Indeed, let $\cA$ be an \NFA such that $\regL(\cA)=(a+b)^*a$ and consider a word $w\in \jL(\cA)$, then $w$  has at least one occurrence of $a$ at some index $i$.
    Then, for the jumping sequence $\vec{a}=(0,1,2,\ldots,i-1,n,i+1,\ldots n-1,i,n+1)$ we have that $w_\vec{a}\in \regL(\cA)$. Observe that $d_H(w_\vec{a},w)\le 2$ (since $w_\vec{a}$ differs from $w$ only in indices $i$ and $n$), and $\turnindices(\vec{a})\subseteq \{i,n\}$, so $\cA_\Ham(w)\le 2$  and  $\cA_\Rev(w)\le 2$.

    For $\Abs$ and $\Max$, however, consider the word $ab^n$ for every $n\in \bbN$. Since the letter $a$ must be read last, any jumping sequence $\vec{a}$ accepting the word has $\vec{a}_n=1$ and $\vec{a}_{n+1}={n+1}$, meaning a jump of length $n-1$ occurs. We therefore have that $\cA_\Abs,\cA_\Max$ are unbounded.
\end{exa}

\begin{exa}
    \label{xmp:UUBB}
    Let $L=((a(aa)^*b(bb)^*)^*+(b(bb)^*a(aa)^*)^*)(a^*+b^*)$. That is, $L$ consists of all the words where every maximal sequence of all $a$ or all $b$ is of odd length, except maybe the last. $L$ is bounded in the $\Rev$ and $\Max$ semantics, but unbounded in $\Abs$ and $\Ham$.
    Indeed, let $\cA$ be an \NFA such that $\regL(\cA)=L$. We have that $\jL(\cA)=\{a+b\}^*$: Let $w\in\{a+b\}^*$. If $w\in(a^*+b^*)$ then clearly $w\in L$. Otherwise, let $m,n$ be the number of occurrences of $a,b$ in $w$ respectively. If either of $m,n$ is odd then $a^mb^n$ or $b^na^m$, respectively, is a permutation of $w$ and is in $\regL(\cA)$. Otherwise, $a^{m-1}bab^{n-1}$ is. To see that $\Max$ is bounded, intuitively, we read the word mostly sequentially, except when reaching the end of a sequence of w.l.o.g $a$'s, we swap the last $a$ of the sequence with the subsequent $b$ if needed to maintain the parity condition.
    Formally, consider the jump sequence $\vec{a}=(0,1,a_2,\ldots a_n,n+1)$ where $a_i$ for $2\le i\le n$ is defined inductively as follows, maintaining the invariant that for all $j$, $a_j\in\{j-1,j,j+1\}$ and $a_j=j-1\iff a_{j-1}=j$ (that is, $\vec{a}$ consists of disjoint transpositions of subsequent indices).     
    We say that $i$ is \emph{post-change} if $w_{i-1}\ne w_i$, and is \emph{pre-change} if $i<n$ and $w_i\ne w_{i+1}$. We consider the following cases:
    \begin{itemize}
        \item If $i$ is neither post-change nor pre-change, then $a_i=i$.
        \item If $i$ is post-change and not pre-change, then $\vec{a}_i=\begin{cases}
            i & a_{i-1}=i-1 \\
            i-1 & a_{i-1}=i \\
            i & a_{i-1}=i-2 \text{ (so $a_{i-2}=i-1$)}
        \end{cases}$
        \item If $i$ is pre-change, then let $j$ be the maximal index such that $j<i$ and $w_{a_j}\ne w_i$, or $j=0$ if such $j$ does not exist. Then $a_i=\begin{cases}
            i & i-j \text{ is odd} \\
            i+1 & i-j \text{ is even}
        \end{cases}$
    \end{itemize}
\noindent 
     It holds that $w_\vec{a}\in\jL(\cA)$. Since $\vec{a}$ consists of disjoint transpositions of subsequent indices we have that $\sem{\vec{a}}_\Max\le 2$ (with a jump of size $2$ possibly occurring if two swaps are made consecutively). In conclusion, $\cA_\Max$ is bounded. By \cref{lem:relationship Max to Rev} we have that $\cA_\Rev$ is also bounded.

    For $\Ham$, however, consider the word $w=(aabb)^n$. For any $k$ and any permutation $w'$ of $w$ with $d_H(w,w')\le k$, $w'$ has at least $n-2k$ sequences of length $2$, and in particular $w'\notin\jL(\cA)$ for a large enough $n$. It follows that $\cA_\Ham$ is unbounded, and by \cref{lem:relationship Abs to Ham}, so is $\cA_\Abs$.
\end{exa}

\begin{exa}
    \label{xmp:UUBU}
    The language $a^*b^*$ is bounded in the $\Rev$ semantics, but unbounded in $\Ham$, $\Abs$ and $\Max$. 
    Indeed, let $\cA$ be an \NFA such that $\regL(\cA)=a^*b^*$ and consider a word $w\in \jL(\cA)$, and denote by $i_1<i_2\ldots<i_k$ the indices of $a$'s in $w$ in increasing order, and by $j_1>j_2>\ldots >j_{n-k}$ the indices of $b$'s in decreasing order. 
    Then, for the jumping sequence $\vec{a}=(i_1,\ldots,i_k,j_1,\ldots j_{n-k},n+1)$ we have that $w_\vec{a}\in \regL(\cA)$, and $\cA_\Rev(w)\le 2$ (since the jumping head goes right reading all the $a$'s, then left reading all the $b$'s, then jumps to $n+1$). 

    For $\Ham$ and $\Max$, consider the word $w=b^na^n$ for every $n\in \bbN$. 
    The only permutation of $w$ that is accepted in $\regL(\cA)$ is $w'=a^nb^n$, and $d_H(w,w')=n$, so $\cA_\Ham$ is unbounded. By~\cref{lem:relationship Abs to Ham} it follows that $\cA_\Abs$ is also unbounded. Additionally, any jumping sequence $\vec{a}$ accepting the word has $\vec{a}_0=0$ and $\vec{a}_1\ge n+1$, thus incurring a jump of length at least $n$, and so $\cA_\Max$ is also unbounded.
\end{exa}

\begin{exa}
    \label{xmp:UUUU}
    The language $(ab)^*$ is unbounded in all the semantics. Indeed, let $\cA$ be an \NFA such that $\regL(\cA)=(ab)^*$.

    Consider the word $w=b^na^n$ for every $n\in \bbN$, and let $\vec{a}=(a_0,a_1,\ldots,a_{2n},a_{2n+1})$ such that $w_\vec{a}\in (ab)^*$, then for every odd $i\le 2n$ we have $a_i\in \{n+1,\ldots, 2n\}$ and for every even $i\le 2n$ we have $a_i\in \{1,\ldots, n\}$. In particular, every index $1\le i\le 2n$ is a turning point, so $\cA_\Rev(w)=2n$, and $\cA_\Rev$ is unbounded. By ~\cref{lem:relationship Ham to Rev,cor:relationship Abs to Rev,lem:relationship Max to Rev}, it follows that $\cA_\Abs,\cA_\Ham,\cA_\Max$ are also unbounded.
\end{exa}

\section{Discussion and Future Work}
\label{sec:discussion}
Quantitative semantics are often defined by externally adding some quantities (e.g., weights) to a finite-state model, usually with the intention of explicitly reasoning about some unbounded domain. 
It is rare and pleasing when quantitative semantics arise naturally from a Boolean model. In this work, we study four such semantics. Curiously, despite the semantics being intuitively unrelated, it turns out that they give rise to interesting interplay (see~\cref{sec:interplay}).

We argue that Boundedness is a fundamental decision problem for the semantics we introduce, as it measures whether one can make do with a certain budget for jumping. An open question left in this research is \emph{existentially-quantified boundedness}: whether there \emph{exists} some bound $k$ for which $\cA_\Sem$ is $k$-bounded. 
This problem seems technically challenging, as in order to establish its decidability, we would need to upper-bound the minimal $k$ for which the automaton is $k$-bounded, if it exists.
The difficulty arises from two fronts: first, standard methods for showing such bounds involve some pumping argument. However, the presence of permutations makes existing techniques inapplicable. We expect that a new toolbox is needed to give such arguments. 
Second, the constructions we present for \probParamBoundedness in the various semantics seem like the natural approach to take. Therefore, a sensible direction for the existential case is to analyze these constructions with a parametric $k$. The systems obtained this way, however, do not fall into (generally) decidable classes.  For example, in the $\Ham$ semantics, using a parametric $k$ we can construct a labelled VASS. But the latter do not admit decidable properties for the corresponding boundedness problem.

We remark on one fragment that can be shown to be decidable: consider a setting where the jumps are restricted to swapping disjoint pairs of adjacent letters, each incurring a cost of 1. 
Then, the \JFA can be translated to a weighted automaton, whose boundedness problem is decidable by~\cite{hashiguchi1982limitedness,leung2004limitedness}. We remark that the latter decidability is a very involved result. This suggests (but by no means proves) that boundedness may be a difficult problem.



\bibliographystyle{alphaurl}
\bibliography{main}

\end{document}